\documentclass{aa}
\usepackage{natbib}

\def\Msun{$M_{\odot}$}
\def\Rsun{$R_{\odot}$}
\def\SB9{$S\!_{B^9}$}
\def\kms{km~s$^{-1}$}
\def\elogP{$(e - \log P)$}

\bibpunct{(}{)}{;}{a}{}{,}

\usepackage[dvips]{graphicx}
\usepackage{times}
\newcommand\ignore[1]{} 
\usepackage{amsmath}

\begin{document}
\title{Spectroscopic binaries among  Hipparcos M giants\thanks{Based on observations carried out at the Swiss telescope installed at the {\it Observatoire de Haute Provence} (OHP, France), and at the 1.93-m OHP telescope}}
\subtitle{III. The eccentricity - period diagram and mass-transfer signatures}
\titlerunning{Spectroscopic Binaries among M Giants}
\author{A. Jorissen
\and A. Frankowski\thanks{Postdoctoral Researcher, F.N.R.S., Belgium.
Currently at Department of Physics, Technion-Israel Institute of
Technology, Haifa 32000, Israel}
\and B. Famaey\thanks{Postdoctoral Researcher, F.N.R.S., Belgium}
\and S. Van Eck\thanks{Research Associate, F.N.R.S., Belgium} 
}
\institute{
Institut d'Astronomie et d'Astrophysique, Universit\'e
Libre de Bruxelles, CP. 226, Boulevard du Triomphe, B-1050
Bruxelles, Belgium   
}

\date{Received date; accepted date} 
\abstract
{This paper is the third one in a series devoted to studying the
  properties of binaries involving M giants.}
{We use a new set of orbits to construct the first $(e - \log P)$ diagram of an extensive sample of 
M giant binaries, to obtain their mass-function distribution, 
and to derive evolutionary constraints for this class of binaries and 
related systems.}
{The orbital properties of binaries involving M giants were analysed and compared
  with those of related families of
binaries (K giants, post-AGB stars, barium stars, Tc-poor S stars).}
{The orbital elements of post-AGB stars and M giants are not very 
  different, which may indicate that, for the considered sample of
  post-AGB binaries, the post-AGB star left the AGB at quite an early
  stage (M4 or so). Neither are the orbital elements of post-mass-transfer
  binaries like barium stars very different from those of M giants,
  suggesting that the mass transfer did not alter the orbital
  elements much, contrary to current belief. Finally, we
  show that binary systems with $e < 0.4 \log P -
  1$ (with periods expressed in days) are predominantly post-mass-transfer systems, because (i) the
  vast majority of barium and S systems match this condition, and (ii)
  these systems have companion masses peaking around 0.6~\Msun, as
  expected for white dwarfs. The latter property has been shown to
  hold as well for open-cluster binaries involving K giants, for which
  a lower bound on the companion mass may easily be set.}
{}
\maketitle
\section{Introduction}
\label{Sect:Intro}

This paper is the third in our series discussing the 
spectroscopic-binary content of a sample of M giants drawn from the 
Hipparcos Catalogue \citep{ESA-1997}, for which CORAVEL radial velocities 
have been obtained in a systematic way \citep{Udry-1997,Famaey-2005}. 
The availability of an extensive set of orbital 
elements for M giants, assembled in Paper~I \citep{Famaey-2008:b}, 
allows us to address here evolutionary issues through the study 
of the eccentricity -- period diagram [denoted $(e - \log P)$ in the following]. 

The comparison in Sect.~\ref{Sect:e-logP} of such 
$(e - \log P)$ diagrams for K and M giants will reveal the 
operation of tidal effects, whereas a comparison of the diagrams of 
pre-mass-transfer systems with post-mass-transfer 
systems (like post-AGB stars in Sect.~2.2 and barium stars in Sect.~2.3) 
will illustrate the impact of the mass-transfer process on the orbital elements. 

Finally, we will show that the impact of mass transfer on orbital elements is 
also readily apparent when considering how mass functions distribute across the $(e - \log P)$ diagram 
(Sect.~\ref{Sect:fM}). This paper presents for the first time compelling evidence 
that the lower-right corner 
of the $(e - \log P)$ diagram contains mostly post-mass-transfer objects 
with presumably white-dwarf companions. This demonstration is based on 
the analysis of the mass functions from K giants in open clusters \citep{Mermilliod-2007b}. 

\section{The $(e - \log P)$ diagram of M giants: Insights into binary evolution}
\label{Sect:e-logP}

\begin{figure}[]
\includegraphics[width=\columnwidth]{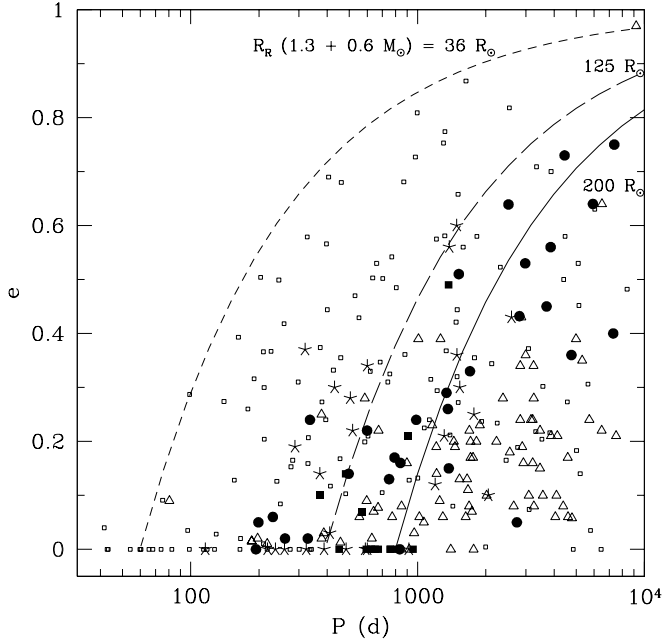}
\caption{\label{Fig:elogP_M}
The \elogP\ diagram for spectroscopic binaries 
involving M giants (black dots,
from this work and collected from the literature), 
K giants in open clusters (small open squares),
barium and S stars (open triangles), symbiotic stars {\em involving an M giant}
(thus excluding yellow symbiotics; filled squares)
 and post-AGB stars (star symbols). 
The short- and long-dashed lines correspond to loci of constant periastron
distance (78 and
280~$R_\odot$), translating into Roche radii of 35 and
125~$R_\odot$, respectively (assuming masses of 1.3 and 0.6 $M_\odot$
for the two components). They provide good fits to the upper envelopes of
the regions where K or M giants with non-circular orbits are located. 
The solid line corresponds to a periastron-distance of 450~$R_\odot$ and a 
Roche radius of 200~$R_\odot$. In Fig.~6 of Paper~II \citep{Frankowski-2008}, 
a different periastron-distance limit was adopted for M giants (70~\Rsun, instead of 125~\Rsun\ here), 
because the important issue in the context of Paper~II was to find a limit enclosing {\it all} M stars, whereas here, a good fit to the envelope of the region occupied by non-circular M giants was seeked.
}
\end{figure}

\begin{figure}[]
\includegraphics[width=1.8\columnwidth]{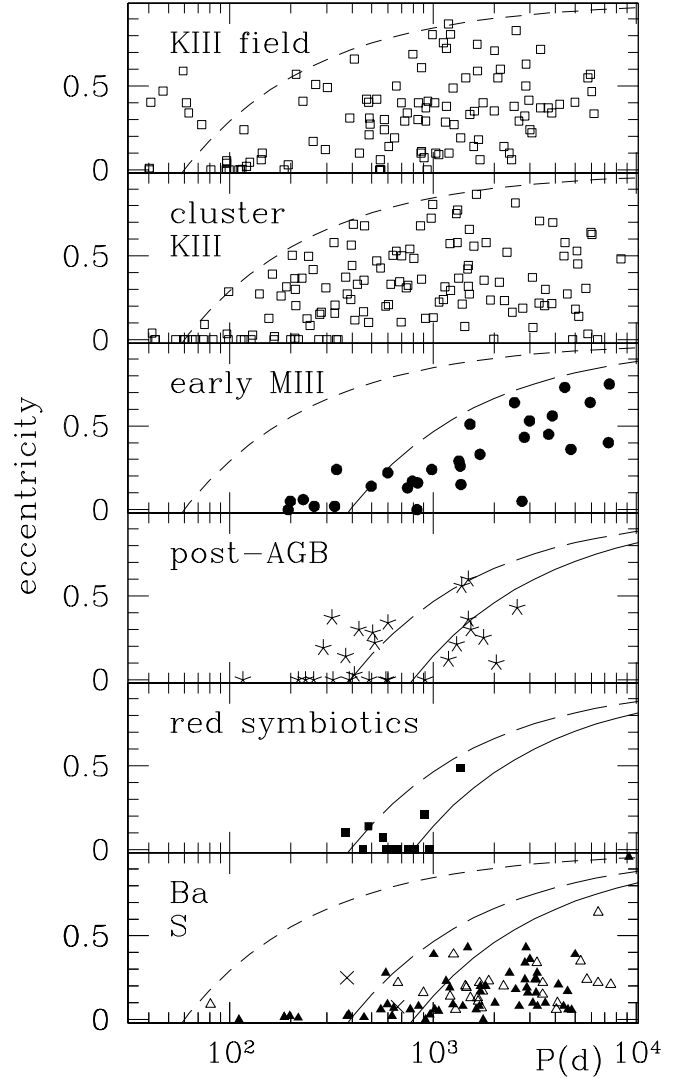}
\caption{\label{Fig:elogP_panels}
Same as Fig.~\ref{Fig:elogP_M} in sequential order, with the
  addition of field K giants from the Ninth Catalogue of Spectroscopic
  Binary Orbits \citep{Pourbaix-04a}, from which the systems with
   sdO/sdB or white dwarf companions (like barium stars) have been
  removed (since they are plotted in the bottom panel).  
For the sake of clarity, all panels start at a period of 30~d,
  although field K giants may be found in binary systems with periods
  as short as 4~d \citep[see Fig.~6 of Paper~II; ][]{Frankowski-2008}.
The lines and symbols are as in
Fig.~\protect\ref{Fig:elogP_M} except for the the S and Ba star panel
where the
crosses denote S stars with an unexpectedly large mass function
(HD~191589 and HDE~332077), open triangles denote mild Barium stars,
and filled triangles strong barium stars and Tc-poor S stars. 
Mild and strong barium
stars are separated according to the photometric DDO index
$\Delta(38-41)$ (see text).   
}
\end{figure}

The $(e - \log P)$ diagram offers good diagnostics about physical
processes at work in binary systems.
The $(e - \log P)$ diagram of M giants, constructed from the new and
literature orbital data collected in  
Table~8 of Paper~I\footnote{ Note that the short-period  orbit of HD~115521, whose Keplerian  nature is in fact questionable (see Sect.~3.3.2 of Paper~I), 
has not been included in the figure.}, is presented in
Fig.~\ref{Fig:elogP_M}. Important clues can be obtained
about the evolution of binaries when comparing the location of M
giants in this diagram with that of other classes of binary stars, namely  
K giants in open clusters \citep{Mermilliod-2007b} or in the
field\footnote{ Systems with
   sdO/sdB or white dwarf companions (like barium stars) have been
  removed from that sample (in order not to duplicate systems between
  the bottom and the top panels
  of Fig.~2).}
\citep[from the Ninth Catalogue of Spectroscopic Binary Orbits;][]{Pourbaix-04a}, barium and
(Tc-poor) S stars 
\citep{Jorissen-VE-98}, symbiotic stars with
M-giant primaries \citep{Belczynski00,Mikolajewska-03,Fekel-2007} and 
post-AGB stars \citep{VanWinckel-2007}.
For the sake of clarity, these different stellar classes are presented
again in separate panels in Fig.~\ref{Fig:elogP_panels}.
In that figure, mild and strong barium stars are separated
according to the DDO index \citep{Lu-91}
$\Delta(38-41) = C(38-41) -0.90\; C(42-45) + 1.33$  being respectively
$\le -0.11$ or $> -0.11$  \citep[see ][ and the Appendix
of Jorissen et al. 1998]{Jorissen-Boffin-92}. 
If the DDO indices $C(38-41)$ and $C(42-45)$ needed to construct the $\Delta(38-41)$
index are not available, the Warner
Ba index has been used instead (respectively Ba $\le 2$ or Ba $\ge 3$
for mild and strong barium stars).  

\begin{figure}
\includegraphics[width=\columnwidth]{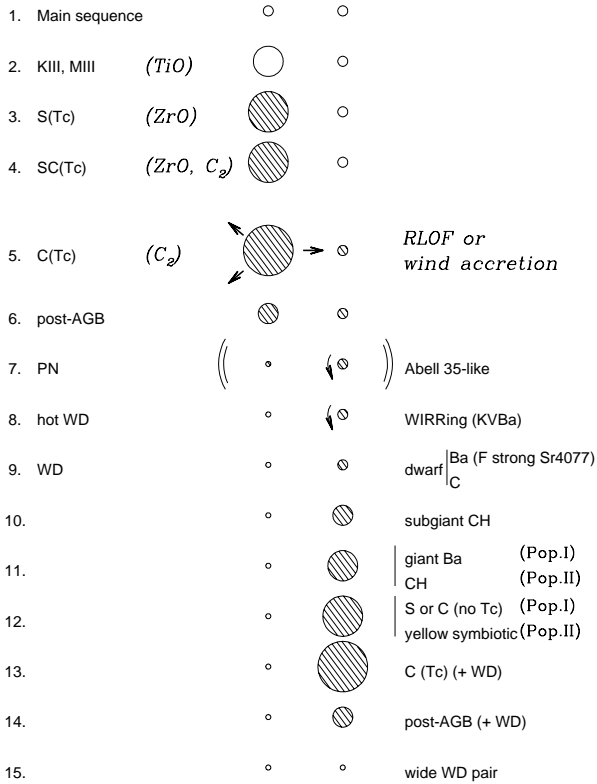}
\caption[]{\label{Fig:binaryscenario}
One among the various possible evolution channels for a
system consisting initially of two low- or intermediate-mass
main-sequence stars. The left column corresponds to the normal
evolutionary sequence of single stars, while the right column
represents the various classes of stars with chemical peculiarities 
specifically produced by mass transfer across the binary system. Hatched circles denote 
stars with atmospheres enriched in carbon or heavy elements \citep[see]
[ for a detailed description of the various stellar families 
involved]{Jorissen-03}. Other evolutionary channels would end
  with a cataclysmic variable or go through an Algol phase, and are
  not depicted here \citep[see, e.g., ][ for
    details]{Nelson-2001,Eggleton-2006}.
}
\end{figure}

The various classes of binary stars presented in
Figs.~\ref{Fig:elogP_M} and \ref{Fig:elogP_panels} are very likely linked through
what may be called the {\em mass transfer scenario} \citep{McClure-83}: the evolution
followed by a system consisting initially of two low- or
intermediate-mass main-sequence stars will go through the various phases  
listed above, along the sequence illustrated in Fig.~\ref{Fig:binaryscenario}.
At point 5 in this evolution, part of the mass lost by the asymptotic giant
branch (AGB) component will be accreted by the companion, thus
increasing the abundances of those elements produced by the AGB nucleosynthesis, most notably
C, F and s-process elements. Note that red symbiotics apparently do not 
exhibit such enhancements and their exact status in this sequence is still 
debated \citep[see][ for recent
  reviews]{Jorissen-03,Jorissen-VanEck-2005,Frankowski-2007a}. 

In some systems (most probably the closest), the binary evolution may branch to an Algol
  configuration at phases 2, 3 or 4 on
  Fig.~\ref{Fig:binaryscenario}. This branching is not discussed here
  \citep[see instead Fig.~10 of ][ and Eggleton 2006]{Jorissen-Mayor-92}. 
In yet other systems, phase 5 of Fig.~\ref{Fig:binaryscenario} may
  give rise to a common envelope, and those systems then end their
  evolution as cataclysmic variables or even coalesce. This situation
  is thought to occur for systems where a component
  with a deep convective envelope fills its Roche lobe 
  \citep{Pastetter-Ritter-89,Eggleton-2006}. This is supposed to have
  occurred in some barium systems, and yet these systems did not end
  up as cataclysmic variables. Why barium and Tc-poor S stars 
did  escape this dramatic fate 
has remained unclear so far \citep[but see ][]{Frankowski-2007a}. 

As binary stars evolve along the sequence displayed in
Fig.~\ref{Fig:binaryscenario}, their orbital elements are expected to
vary, be it due to tidal effects, to interaction with a circumbinary
disc, or to mass loss or mass transfer \citep[through wind accretion or Roche Lobe
  overflow -- RLOF; see][ for recent discussions]{Eggleton-2006,Frankowski-2007a}.  
Therefore, the comparison of orbital
elements for binary stars located at different stages in the sequence
is expected to shed light on the various physical processes that
have occurred along their evolution. The availability of the extensive
set of orbital elements for M giants obtained in the present study 
brings an important element to this comparison. 

\subsection{Smooth evolution KIII -- MIII -- Ba/S}
\label{Sect:evolution}

Figures~\ref{Fig:elogP_M} and \ref{Fig:elogP_panels} reveal a smooth
evolution along the sequence KIII -- MIII -- Ba/S, in
the sense that the upper boundary of the populated region in the $(e -
\log P)$ diagram
moves towards longer periods (this is reflected by the three curved
lines which roughly delineate the regions populated by these three
classes, their exact definition being given below). 
This is clearly a consequence of the larger radii
reached by stars evolving along this sequence. In the case of Ba and
Tc-poor S giants,
it is actually their white dwarf (WD) companions which reached very large radii
while evolving on the AGB. For K giants, the situation is 
in principle somewhat more complex, since this class mixes stars on the first
giant branch and stars in the core He-burning phase. \textit{Low-mass} stars belonging to the latter category have gone through the RGB tip, where they reached a very large radius (similar to, or even larger than that of M giants). 
Therefore, if those low-mass, core-He burning stars were to dominate among K giants, their distribution in the $(e - \log P)$ diagram should be characterised by an envelope located at even longer periods than that for M giants.  Fig.~\ref{Fig:elogP_panels} shows that this is not the case, because the sample of open-cluster K giants plotted in  Fig.~\ref{Fig:elogP_panels} is in fact dominated by intermediate-mass stars, as may be judged from the turnoff masses of the corresponding clusters, most of them being larger than 2~\Msun\ \citep{Mermilliod-2007b}.   The complication introduced by the mixture of evolutionary states among K giants 
is thus not a concern. 

Equating the stellar radius to the 
Roche radius results in a threshold period (for given component
masses) below which the primary star
undergoes RLOF. Adopting Paczy\'nski's usual expression
for the Roche radius $R_{R,1}$ around star 1
\begin{equation}
\label{Eq:Roche}
R_{R,1}/A = 0.38 + 0.2 \log q \:\;\;\;\;\;\;(0.5 \le q  \le
20),
\end{equation}
where $q = M_1/M_2$ and
$A$ is the orbital separation, one finds that a star of radius
40~R$_\odot$ fills its Roche lobe in a system of period $P = 70$~d,
for masses $M_1 = 1.3 $~M$_\odot$ and $M_2 =
0.6$~M$_\odot$. Although the Roche lobe
concept is in principle only applicable to circular orbits, one may
formally compute the orbital periods for which the primary star fills
its Roche lobe {\it at periastron}, by replacing $A$ by $A(1-e)$ in the
above expression. It is
quite remarkable that the relationship between $P$ and $e$ so obtained
(assuming $R_R = 36$~R$_\odot$) exactly matches the boundary of the
region occupied by KIII giants in the $(e -  \log P)$ diagram, both
for cluster and 
field\footnote{ In the sample of KIII binaries in the field, there
    are several systems with non-zero eccentricities falling to the
    left of the periastron envelope. Most of these systems are either pre-main
    sequence (PMS) binaries \citep[mostly from ][]{Torres-2002} or
    Algols \citep{Nelson-2001,Eggleton-2006}, i.e., (post-)mass-transfer
    systems (see also the discussion in relation with Fig.~6 of
    Paper~II). The successive panels in Fig.~\ref {Fig:elogP_panels}
are meant to illustrate the evolution of the orbital elements as the
system evolves from both components on the main sequence to one
component being a white dwarf. Therefore, the Algols or PMS systems
present in the sample of field KIII stars distort the picture, and
should actually be disregarded.  
} giants (Fig.~\ref{Fig:elogP_panels}). 
This excellent match thus clearly suggests that  mass
transfer at periastron plays a crucial role in shaping the
$(e -  \log P)$ diagram 
\citep{Soker00}.

It may seem  surprising that the 'periastron envelope' [$A(1-e)$
    = constant, or $P^{2/3}(1-e)$ = constant] represents a better fit
  to the data than the 'circularisation envelope' [$A(1-e^2)$ = constant,
    or $P^{2/3}(1-e^2)$ = constant; see Fig.~6 of Paper~II],
  resulting from the fact that circularisation keeps the angular
  momentum per unit reduced mass constant
  \citep{Zahn-1977,Hut81,Duquennoy-92}. Indeed, as the star gets
  closer to its Roche lobe, it should circularise first and then fill
  its Roche lobe, and possibly disappear from the sample due to
  cataclysmic mass transfer. The samples of K-giant binaries clearly
  favour the periastron envelope over the circularisation envelope.
  The reason for this may be the following. When a system is close to
  filling its Roche lobe at periastron, circularisation proceeds 
fast at this moment. But since the circularisation path in 
the $(e - \log P)$ diagram is steeper than the periastron line (see Fig.~6 of Paper~II),
circularisation drags the system away from Roche lobe filling at periastron
and thus also slows down its own rate.
Circularisation can accelerate only when other effects bring the star
and its Roche lobe closer again. Therefore an ensemble of binaries traces
on their $e$ -- $\log P$ diagram the current periastron envelope rather  
than any circularisation path (see also Paper~II, Sect.~2.5).

The role played by periastron RLOF in shaping the \elogP\ diagram is nicely
illustrated by the M4III star HD~41511 (= SS~Lep). With $P = 260.3$~d and $e
= 0.02$ \citep{Welty-1995}, it is the fourth shortest orbital period among the
spectroscopic systems involving an M giant. \citet{Verhoelst-2007}
provided convincing evidence that the M giant, having a radius of
$110\pm30$~\Rsun\ derived from interferometric measurements and from the
Hipparcos parallax (Table~\ref{Tab:radius}), fills its Roche lobe. 
This is consistent with the Roche limit set at 125~\Rsun\ in Figs.~\ref{Fig:elogP_M} and \ref{Fig:elogP_panels}.
As a further evidence to
this claim, these authors show that the A-type companion appears to be
a main-sequence star, which has swollen as a result of accretion from the
lobe-filling M giant. Indeed, its radius is ten times that of a
typical A1V star, but its mass, derived from the surface gravity and
radius, is not that of a massive supergiant. Finally, the system is
surrounded by a circumbinary disc \citep{Jura-2001,Verhoelst-2007},
probably fed by the RLOF, which
thus appears to be non conservative.   
The A star is rapidly rotating \citep[$V\sin i =
  117$~\kms;][]{Royer-2002}, possibly as a result of spin accretion or, alternatively,
as the primordial rotation speed of the main sequence star. All these features are very suggestive of a scenario outlined by 
\citet{Frankowski-2007a} for an evolving M giant binary.

For M giants with radii somewhat smaller than that of SS Lep, but
still representing a fair fraction of the Roche radius, tidal effects
will take place, and the star will then be seen as an ellipsoidal
variable.  The M2.5III star
HD~190658 (V1472~Aql) is such a case \citep{Samus-1997}. It has the second shortest orbital period among M
giants \citep[$P = 199$~d, $e = 0.05$; ][]{Lucke-Mayor-1982}.
A rotational velocity of $V \sin i = 15.1$~\kms\ is found from the CORAVEL cross-correlation dip width calibrated by \citet{Benz-1981}.  
Assuming that the giant star rotates in synchronism with the orbital
motion, a radius of 59~\Rsun/$\sin i$ is derived, corresponding to
$R/A_1 = 1.16$, where $A_1$ is the semi-major axis of the giant's
orbit around the centre of mass of the system
\citep{Lucke-Mayor-1982}. Adopting typical masses of 1.7~\Msun\ for
the giant and 1.0~\Msun\ for the companion, one obtains $R/A = 0.43$
and $R/R_R = 1$ ! 

Interestingly, the Ba/S star HD~121447 (with an
effective temperature equivalent to that of a M0III giant) is another
example: with $P = 185.7$~d and $e = 0$, it falls just besides
HD~190658 in the $(e - \log P)$ diagram and seems to be an ellipsoidal variable
as well \citep{Jorissen-95}, although that interpretation has been
challenged by \citet{Adelman-2007}, based on the presence of chromatic
variations. 

HD~9053 (=$\gamma$~Phe), with $P= 193.8$~d and $e = 0$,  has the shortest period among M-giant binaries.
\citet{Jancart-2005} used the spectroscopic elements and the Hipparcos data to derive a combined orbit yielding a precise inclination $i = 46\pm4^\circ$. The system cannot therefore be eclipsing, so that  the eclipsing-like light curve reported by \citet{Otero-2007} should in fact be re-interpreted in terms of ellipsoidal variations, a possibility mentioned by \citet{Otero-2007}. The angular radius measured by \citet{Richichi-2005b} is $3.22\pm0.43$~mas, yielding 50~\Rsun\ when combined with the Hipparcos parallax of 
$13.94\pm0.64$~mas, to be compared with a Roche radius of 78~\Rsun\ (adopting masses of 1.3 and 0.6~\Msun\ for the giant and its companion, respectively). The giant should therefore be well within its Roche lobe. Nevertheless,  HD~9053 is a 'hybrid star'  \citep[][]{Ayres-2005}: despite being 
located to the right of the 'dividing line' between hot coronae and massive winds \citep{Hunsch-1996,Reimers-1996}, the star exhibits signatures of hot gas, like X-ray flux and C~IV line.  One may wonder whether these could be signatures of matter being accreted by the companion in this close binary system. Also note that arguments were presented in Sect.~2.5 of Paper~II that the usual expressions for the Roche radius \citep[e.g.,][]{Eggleton-83} may actually overstimate it when an extra-force (like radiation pressure) is present in the system \citep[see also ][]{Dermine-Jorissen-2008}.

We will come back to the
relationship between ellipsoidal variables and the periastron envelope
observed in the  $(e - \log P)$ diagram when we will show in
Sect.~\ref{Sect:D} that the so-called long secondary periods (also
called sequence D) in the $K - \log P$ diagram of long-period variables (LPVs)  in the Large
Magellanic Cloud \citep{Wood-1999,Wood-2000} may be identified with the
periastron envelope in the $(e - \log P)$ diagram of galactic M giants
\citep[see also ][]{Soszynski-2007}.   

No similar evidence for RLOF or ellipsoidal variations has been found for
the other  two short-period M
giants (HD~89758 and HD~147395).

\begin{figure}
\includegraphics[width=\columnwidth]{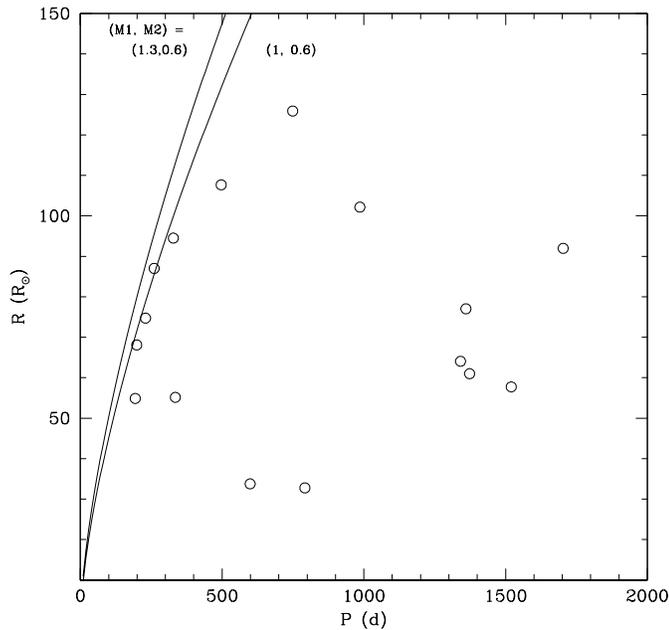}
\caption[]{\label{Fig:radius}
The orbital-period -- radius diagram for M giants in binary
systems. The radii are derived from the Stefan-Boltzmann law applied
on the effective temperatures and luminosities presented in
Fig.~\ref{Fig:HR}. The solid lines display the Roche radii for the
corresponding period, for two different choices of the component masses. 
}
\end{figure}

The periastron-distance envelope seen in the \elogP\ diagram
(Fig.~\ref{Fig:elogP_panels}) is also apparent in the radius --
orbital-period diagram (Fig.~\ref{Fig:radius}).
The radii plotted in Fig.~\ref{Fig:radius} are derived from the
Stefan-Boltzmann law and  
are in very good agreement with those derived by combining the 
uniform-disc diameters listed in the 'updated Catalogue of High
Angular Resolution Measurements' \citep[CHARM2;][]{Richichi-2005} and the
Hipparcos parallaxes (Table~\ref{Tab:radius}).
 Fig.~\ref{Fig:radius} shows that, for a given value of the
radius,  there is a lower bound on the orbital
period\footnote{\citet{Pourbaix-04a} have presented a spectacular
  figure (their Fig.~4) showing the minimum orbital period as a function of the $B-V$ index 
for the whole {\em Ninth Catalogue of Spectroscopic Binary Orbits}
(SB9).}, simply because stars with large
radii cannot reside in too close systems, or they would undergo RLOF
and would disappear from the sample by transforming into
cataclysmic variables.  
The lines in Fig.~\ref{Fig:radius} correspond to the limit imposed by equating
the stellar radius to the Roche
radius (Eq.~\ref{Eq:Roche}, adopting $M_1 = 1.3$~\Msun\ and $M_2 =
0.6$~\Msun, or  $M_1 = 1.0$~\Msun\ and $M_2 =
0.6$~\Msun), 
the stellar radius being computed using
the Stefan-Boltzmann relation as explained before. 

\begin{table*}
\caption[]{\label{Tab:radius}
Radii of M giants in binary systems derived from the Stefan-Boltzmann
law ($R_{\rm SB}$) and from high-angular resolution measurements ($R_{\rm HR}$).
 The labels
'UD' and 'LD' refer to uniform discs or limb-darkened discs, respectively.
Luminosities are derived from the
Hipparcos parallaxes $\varpi$, from the 2MASS $K$ magnitudes and bolometric corrections
$BC_K$ derived from the ($V-K$ , $BC_K$) analytical relation from 
\citet{Bessell-Wood-84}, which is in good agreement with
the more recent calibration of \citet{Bessell-98}. The Johnson $V_J$
magnitude from the Hipparcos catalogue has been used. Effective
temperatures are derived from the $V_J-K$ indices using the 'bcd' calibration of \citet{Bessell-98}. 
}
\begin{tabular}{rrcrrrrrlllll}
\hline
\hline
\multicolumn{1}{c}{HD} & \multicolumn{1}{c}{$\theta/2$} & UD/LD &
\multicolumn{1}{c}{band} & \multicolumn{1}{c}{$\varpi$} & $R_{\rm HR}^a$
  & $R_{\rm SB}$ & \multicolumn{1}{c}{$K$ (2MASS)} & \multicolumn{1}{c}{$V_J-K$} & \multicolumn{1}{c}{$BC_K$}  & \multicolumn{1}{c}{$\log L/L_\odot$} & $\log T_{\rm eff}$ \\
   &  \multicolumn{1}{c}{(mas)}      &       &      & \multicolumn{1}{c}{(mas)}    & ($R_\odot$) &
($R_\odot$) & & & & & (K)\\
\hline
9053  & $3.22\pm0.43$ & UD & 10 $\mu$m & $13.9\pm0.6$ & 50 & 55 & -0.50$^b$ & 3.9 & 2.65 & 2.75 & 3.58\\
41511 &  $1.55\pm0.16$  & LD & K & $3.0\pm0.7$ & 110 & 109 &  1.67 & 3.3 &
- & 3.08$^c$ & 3.51$^c$ \\ 
42995 &  $12.95\pm0.04$ & UD & K & $9.3\pm2.0$ & 145 & 153 & -1.72 & 5.0 &
2.85 & 3.50 & 3.55 \\ 
89758 &  $ 8.00\pm0.03$ & UD & V &$13.1\pm0.8$ &  66 &  75 & -1.01 & 4.1 &
2.69 & 2.99 & 3.57 \\
132813 & $ 9.60\pm0.70$ & UD & K & $8.2\pm0.5$ & 126 & 126 & -0.96 & 5.6 &
2.91 & 3.29 & 3.54 \\
137853 & $ 2.37\pm0.05$ & UD & K & $4.2\pm0.8$ &  62 &  61 &  1.93 & 4.1 &
2.69 & 2.81 & 3.57 \\ 
220088 & $ 2.29\pm0.03$ & UD & K & $7.5\pm0.7$ &  33 &  35 &  1.81 & 3.8 &
2.61 & 2.38 & 3.59 \\  
\hline
\end{tabular}

$^a$ From Richichi et al. (2005; CHARM2) \nocite{Richichi-2005}, 
\citet{Verhoelst-2007} and \citet{Richichi-2005b} \\
$^b$ From \citet{Richichi-2005b}\\
$^c$ These values are taken from \citet{Verhoelst-2007}; the
calibrations used for the other stars, and based on the $V-K$ colour
index, cannot be used for HD~41511 because the $V$ band is
contaminated by light from the A-type companion
\end{table*}

Finally, it is noteworthy that the solid line in
Figs.~\ref{Fig:elogP_M} and \ref{Fig:elogP_panels} does not match
well the upper boundary of the region occupied by barium and Tc-poor S stars,
thus suggesting that for these classes of post-mass-transfer binaries,
tidal processes and periastron mass-transfer were, as expected, not the only ones to operate. 
Mass transfer and interaction with a circumbinary disc must have
played a role for these \citep{Frankowski-2007a}. One mild barium star
(HD~77247: $P = 80.5$~d; $e = 0.09$) deviates markedly from the rest of the sample of
barium stars. One may
wonder whether the location of HD~77247 in the \elogP\ diagram    
along the periastron envelope of K giants is just accidental, or whether it
indicates instead that this system might not be a post-mass-transfer
system after all, as are the other barium stars, but that its barium anomaly may
rather be inherited from the gas out of which it formed, and be unrelated to its binary nature. HD~77247 would thus join the small number of (mostly dwarf) barium stars with no evidence for being binaries \citep[see the review by][ and individual cases discussed by Tomkin et al. 1989, Edvardsson et al. 1993, North et al., 1992, 2000]{Jorissen-03}. \nocite{Tomkin-89,Edvardsson-93,North-92,North-00}

\subsection{Relation with post-AGB stars}
\label{Sect:pAGB}

Binary post-AGB stars \citep{VanWinckel-03,VanWinckel-2007} are the immediate
descendants of the binary M giants (Fig.~\ref{Fig:binaryscenario}), and indeed 
they share almost exactly the same location in the $(e - \log P)$ diagram. 

The M4III + A system SS~Lep already mentioned in Sect.~\ref{Sect:evolution} is
very interesting in this respect, since it shares with post-AGB stars the presence of a circumbinary disc and its position in a group of binaries with the shortest orbital periods among both families.
It may thus be considered as a system linking binary M giants
and post-AGB systems in the sense that the M-giant component in SS~Lep should shortly evolve into a post-AGB star.

On the contrary, one may wonder whether some among the post-AGB stars from panel 4 in Fig.~\ref{Fig:elogP_panels} with circular orbits and $P \sim 200$~d like SS~Lep (these are  HD~101584, HD~213985 and 
IRAS~05208-2035, although in the case of HD~101584, the eccentricity is not well constrained) would not be genuine post-AGB stars, but rather be 
accreting main-sequence or white-dwarf stars which  have swollen to
giant dimensions in RLOF systems \citep[as expected when the accreting
  main-sequence star has a
radiative envelope:] [ or when the WD accretion rate is large enough to
  trigger shell H-burning: Paczy\'nski \& Rudak 1980, Iben \&
  Tutukov 1996]{Kippenhahn77,Jorissen-03}. The post-AGB stars in those systems would thus be the analogs of the A-type companion of SS~Lep. \nocite{Paczynski-Rudak-80,Iben-Tutukov-1996} 

The spectrum of HD~101584 is very complex, combining emission lines with P-Cygni profiles indicative of ongoing mass loss and absorption lines. The remarkable characteristic of the optical spectrum of 
HD~101584 is the fact that different spectral regions resemble different spectral types. There is therefore no agreement about the effective temperature of the central star, with proposed values ranging from 8500~K \citep{Sivarani-1999} to 12\ts000~K \citep{Bakker-1996}, or $8500\pm1000$~K for the post-AGB star and about 11\ts000~K for its companion \citep{Kipper-2005}. There is thus no clear evidence from recent works for the companion of the post-AGB star HD101584 to be an M giant, despite the early suggestion by \citet{Humphreys-1976}.

\begin{figure}
\includegraphics[width=\columnwidth]{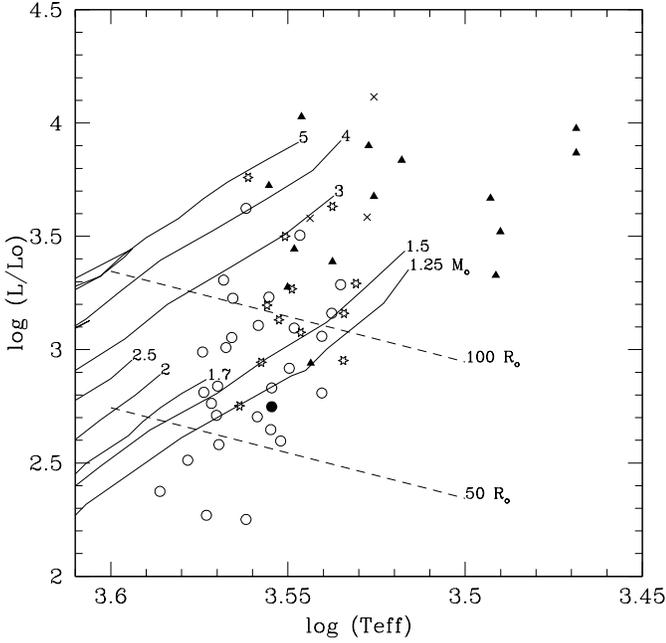}
\caption[]{\label{Fig:HR}
Comparison of the location of M giants in binary systems (open circles) and Tc-rich S
stars (filled triangles) in the HR diagram. The filled circle
corresponds to the M giant HD~165374 which possibly hosts a WD
companion. Star symbols correspond to Tc-poor S stars and crosses to S
stars with Tc doubtful. The data for S
stars are from \citet{VanEck-98}. For M stars, they are derived as explained in the caption of Table~\ref{Tab:radius}.  The evolutionary tracks are from
\citet{Charbonnel-1996} for $Z=0.02$ stars with masses 1.25 and 1.5~\Msun, and from
\citet{Schaller-1992} for the more massive stars. In all cases, the
tracks span the early AGB and stop at the first thermal pulse.
}
\end{figure}

In the spectral energy distributions presented by \citet{DeRuyter-2006} for HD~213985 and 
IRAS~05208-2035, neither is there evidence for the presence of a cool giant
companion, as it was the case for SS~Lep
\citep{Verhoelst-2007}. 
\renewcommand{\thefootnote}{\alph{footnote}}
Nevertheless, if the luminosity ratio between the
post-AGB and the giant components is of the order 5 to 10\footnote{\citet{DeRuyter-2006} suggest that $\log L/L_{\odot} =
  3.6$ may be typical for post-AGB stars, and Fig.~\ref{Fig:HR}
  reveals that M giants may be as dim as $\log L/L_{\odot} =
  2.6$.}, one may wonder whether the cool giant could at all be
noticed (fitting the complex spectral energy distributions of post-AGB systems is a delicate operation involving many free parameters --like the reddening-- with no unique solution; Van Winckel, priv. comm.). 
\renewcommand{\thefootnote}{\arabic{footnote}}
For the SS~Lep system, the luminosity ratio is
1.6 in favour of the A component \citep{Verhoelst-2007}.
HD~172481    is another example of a F2Ia
supergiant traditionally considered to be a post-AGB star, but which
is more likely a white dwarf burning hydrogen accreted from its AGB
companion \citep{Whitelock-2001}. 
For that system, the presence of
an M giant companion is apparent from the detection of TiO bands, from the
SB2 nature of the CORAVEL cross-correlation dip \citep{Reyniers-2001},
and from Mira-type
pulsations with a period of 312~d. From the SED, the
luminosity ratio turns out to be  similar to SS Lep: $L_{\rm F}/L_{\rm
  M} = 1.8$
(although that ratio is somewhat dependent upon the adopted
reddening). No radial-velocity
variations hinting at a period of a few hundred days have been
detected for HD~172481 \citep{Reyniers-2001}.

The papers mentioned above suggest different types of accreting
components in SS~Lep and HD~172481 but, in the absence of
stringent constraints on the mass and luminosity of the accreting
star, there is in fact no easy way
to distinguish  between WD and main-sequence accretors. 

In summary, we conclude that some systems currently flagged as hosting a post-AGB star could in fact be better described as systems where an M giant dumps matter on its white-dwarf or main-sequence companion which then acquires supergiant dimensions, thus mimicking a post-AGB star. Such systems, currently plotted in panel 3 of 
Fig.~\ref{Fig:elogP_panels} (post-AGB systems), do in fact belong to panel 2 (M giants), were it not for the light of the M giant being lost in the glow of its supergiant-like companion. In this scenario, one might however worry whether the accreting component could indeed become more luminous than the moss-losing giant.

Another possible scenario to account for systems like HD~172481  could therefore be as follows, now stating instead that the post-AGB star is genuine, but its giant companion is fake. In a system with components of initial masses $(M_1,M_2) = (3,1)$~\Msun, the star of mass $M_1 > M_2$ evolves faster, fills its Roche lobe on the AGB around the first thermal pulse and becomes a white dwarf (WD) with a mass of 0.55~\Msun\ \citep{Bloecker-95}. It is not  exactly clear what happens to the orbit, 
although \citet{Frankowski-2007a} presented arguments for the orbital period remaining
more or less unchanged (see also the discussion in Sect.~\ref{Sect:Ba}). Then the star of mass $M_2$ (which could by now have reached   2--3~\Msun\ as a result of accretion) 
evolves in turn along the AGB, but having a lighter companion it can grow bigger and brighter 
without filling its Roche lobe, say up to a core mass of 0.6~\Msun. Eventually star $M_2$ fills its Roche lobe 
and ejects most of its envelope, transferring part of it back to star $M_1$ \citep[a few 
0.01~\Msun\ is enough for $M_1$ to reignite its H shell and to swell again to giant dimensions;][]{Frankowski-2003}. 
There are now two cores with comparable envelope masses, both burning hydrogen
in shells, the younger ($M_2$, i.e. the true-post AGB) burning it faster (as it has a more massive core). The younger core is brighter and evolves (contracts) faster. 
The difference in evolutionary speed becomes enormous as star $M_2$ gets to its 
post-AGB stage \citep[3000 years off the AGB, the 0.6~\Msun\ core has already a temperature of $\log 
T_{\mathrm eff} = 4.5$;][]{Bloecker-95} while star $M_1$ still sits on the AGB. Soon we are 
left with a giant configuration around $M_1$'s core and a hot, brighter 
post-AGB $M_2$. In this particular case the brightness difference is 
a factor of 3 to 5. 
The mass function would be high (0.16~\Msun\ for the sample case above, 
assuming $e = 0, i = 90^\circ$ -- which makes it very similar to SS~Lep, which has $f(M) = 0.26$~\Msun). 

In any case, the suggestion that some stars flagged as post-AGB could in fact be
components of binary systems caught in the act of mass transfer     
would have several interesting consequences. It could explain 
the unexpectedly large number of post-AGB systems (as compared to
their short evolution time scale), the similarity between the orbital
elements of post-AGB and M binaries (if post-AGB stars were {\it
  post}-mass-transfer systems rather than systems with {\it on-going} mass
transfer as we suggest, the orbital elements should have been altered, at least to
some extent). It would also explain why we could not find any other
systems with characteristics similar to SS~Lep [namely, early-type
star with cool dust in a disc and spectral features typical of (super)giants],
since they would all already have been flagged as post-AGB
systems! The star 3~Pup (HD~62623) is sometimes presented as the
twin of SS~Lep \citep{Jura-2001}, but the exact nature of this star is
still much debated. At this point, it should be mentioned that the d' symbiotic systems could be related as well to post-AGB systems and SS~Lep: they host a warm, fast-rotating  giant (F to early K) with a  hot compact companion which just left the AGB, as suggested by the presence of cool dust and, often, a planetary nebula. The fast rotation of the giant is very likely the signature of a recent mass-transfer episode \citep{Jorissen-Zacs-2005,Zamanov-2006}.

\begin{figure}
\includegraphics[width=\columnwidth]{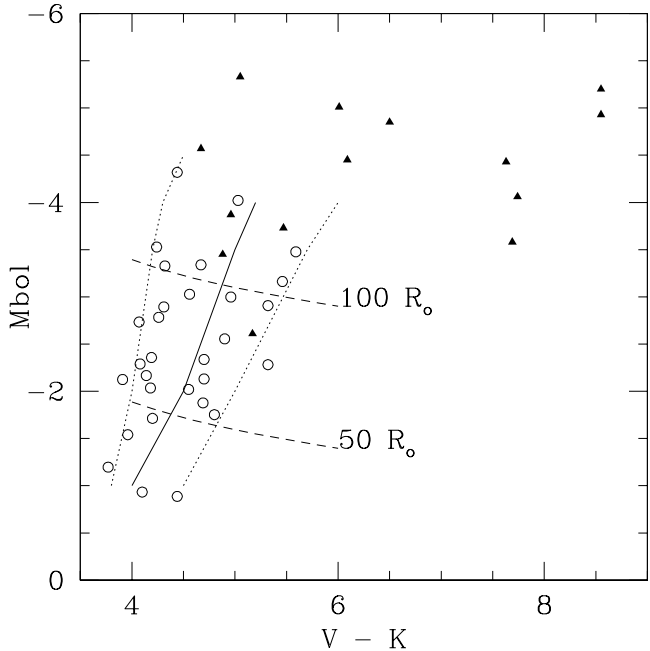}
\caption[]{\label{Fig:HR1}
Same as Fig.~\ref{Fig:HR} but in terms of $V-K$ and $M_{\rm bol}$. The
dotted lines delineate the region occupied by M giants. They
define the range of possible stellar radii for a given $M_{\rm bol}$,
used to construct the lower period bound from the Roche limit in
Fig.~\ref{Fig:MbolP}.  
}
\end{figure}

Actually, the shape of the \elogP\ diagram of post-AGB stars, when
compared to that of binary M and K giants, may be interpreted in two
different ways.
The first interpretation assumes that 
the location of post-AGB stars in the $(e - \log P)$ diagram is
controlled by the periastron envelope corresponding to 200~\Rsun\ (solid
line in Fig.~\ref{Fig:elogP_panels}), with some circular systems at
shorter periods {\em plus} some among these which had their
eccentricities pumped upwards as a result of the interaction with the
circumbinary disc \citep{Artymowicz91,DeRuyter-2006,Frankowski-2008:b}.  In this
scenario, the AGB precursors were thus allowed to evolve far up the
AGB, up to at least 200~\Rsun. The similarity
between the \elogP\ diagrams of M giants and post-AGB systems would
thus be purely accidental.

The second interpretation assumes that this similarity is not
accidental, and states that mass transfer or disc interaction has not dramatically
altered the location of post-AGB binaries in the $(e - \log P)$ diagram
(or as suggested above, some might even be {\em in the course} of the
mass-transfer process). This case then differs from the
first interpretation by implying that the AGB precursors 
must have left the AGB at a rather early
stage in their evolution, 
given their location in the $(e - \log P)$ diagram between the tidal
envelopes corresponding to 85 and 200~\Rsun\ 
(Fig.~\ref{Fig:elogP_M}). 
Like the M giants, {\em binary} post-AGB stars do not show evidence for s-process
enrichment \citep[despite the confusion introduced by the depletion
  pattern;][]{VanWinckel-2007}. We believe that the absence of
s-process enrichments in both classes of stars is not a coincidence.
In fact, it may be inferred that  neither M giants nor post-AGB binaries reached a point on the AGB where 
s-process and dredge-ups were operating\footnote{Among
the 4 MIII stars checked for the presence/absence of Tc \citep[HD~2411 =
  TV~Psc; HD~132813 = RR~UMi; HD 187076 = $\delta$~Sge; HD~42995 =
  $\eta$~Gem;][]{Little-87}, none has been flagged as exhibiting Tc
lines (only in HD~2411 is the presence of Tc qualified as doubtful;
note that this M3III star falls beyond the 200~\Rsun\ tidal envelope in Figs.~\ref{Fig:elogP_M} and \ref{Fig:elogP_panels},
i.e., in the region occupied by barium stars).}.  
This argument may actually be checked by locating M giants and post-AGB stars 
in the
Hertzsprung-Russell (HR) diagram (Figs.~\ref{Fig:HR} and \ref{Fig:HR1}) and comparing their
location with that of Tc-rich S stars \citep[from][]{VanEck-98}, 
which are thermally-pulsing AGB (TP-AGB)
stars enriched in s-process elements. 
Luminosities of post-AGB stars are extremely scarce. 
When available \citep[e.g., from LMC objects and from a
  luminosity-period relationship for RV Tau stars calibrated on LMC
  stars;][]{Alcock-1998,DeRuyter-2006} however, they are of the order
of 4000~L$_\odot$ ($\log L/{\rm L}_\odot = 3.6$). This luminosity lies  in the middle of the range covered by Tc-rich S stars (Fig.~\ref{Fig:HR}),  so that this argument does not really support a non-TP-AGB origin for post-AGB stars. However, the argument strongly relies on the adopted (uncertain) value for the typical  luminosity of post-AGB stars.

\subsection{Relation with S and barium stars}
\label{Sect:Ba}

We now compare M giants in binaries with Tc-rich S stars.
As expected, M giants are always
less evolved than Tc-rich S stars. As already discussed by
\citet{VanEck-98}, Tc-rich S stars are indeed found beyond the onset
of the TP-AGB (i.e., to the right of the early AGB evolutionary tracks
in Fig.~\ref{Fig:HR}). There is one exception, though [HIP 38502 =
NQ~Pup, a Tc-rich S star with $\log(L/L_\odot) = 2.9$], 
which illustrates the possible impact of biases when
considering the individual locations of stars with non-negligible
relative errors on their parallaxes \citep[see][ and references therein]{VanEck-98}.     
Nevertheless, overall, the segregation between Tc-rich S stars and M
giants is very clear. 

When M giants evolve further on the AGB, they will indeed turn into Tc-rich S
stars, but at the same time are likely to become long-period variables
\citep[LPVs;][]{Little-87}.  As discussed in Papers~I and II, 
binary systems among LPVs are very difficult to detect for two
reasons. First, given their large
radii, only long-period systems may host them (see
Fig.~\ref{Fig:radius}), but these systems necessarily have small
values for the radial-velocity semi-amplitude $K$ \citep[smaller than
  about 10~\kms; see Fig.~5 in Paper~II and the
discussion about Fig.~9.4 in][]{Jorissen-03}. Secondly, since they are
LPVs, shock waves move across their atmospheres, and induce
radial-velocity variations with amplitudes of
10 to 20~\kms\ \citep{Hinkle-1997,Alvarez-2001}. Hence they are very
difficult to detect. Spectroscopic binaries involving LPVs are not included in the present
discussion since very few cases are known, and usually their orbital
elements are not known (the supposed post-AGB + Mira system HD~172481 discussed
in Sect.~\ref{Sect:pAGB} is no exception). The list of AGB stars with composite
spectra compiled by \citet{Jorissen-03} includes for instance the
Tc-rich S stars W~Aql, WY~Cas and T~Sgr, and the carbon stars SZ~Sgr,
TU~Tau and BD~-26$^\circ$2983, but orbital periods
are not known for them yet. Symbiotic stars hosting a Mira variable
\citep[the so-called d-type symbiotics;][]{Allen-82,Whitelock87} are perfect examples of binaries
involving a very evolved AGB star. \citet{Schmid-Schild-2002} have
succeeded in detecting the orbital motion of these systems using Raman
polarimetry, and concluded that the orbital periods in these systems
are of the order of 150~yr or 55\ts000~d. 
Intringuingly,
\citet{Goldin-Makarov-2007} find a very short period of 141~d (from
a fit to the Hipparcos Intermediate Astrometric Data) for the M5III
semiregular variable star HIP~34922 (=HD~56096 = L$^2$~Pup), possibly
a TP-AGB star since \citet{Little-87} find Tc to be possibly
present. The period -- eccentricity pair obtained by
\citet{Goldin-Makarov-2007} $(P = 141\pm2$~d, $e =
0.52\pm{0.30\atop0.21})$ is, however, totally inconsistent with the
values found for the other M giants in Fig.~\ref{Fig:elogP_panels}, 
so that its validity must be questioned, as discussed in the Appendix.  

It must be emphasised at this point that {\em Tc-poor} S
stars\footnote{Tc-poor S stars owe their chemical peculiarities to
  mass transfer in a binary system; they are thus called {\em
    extrinsic} S stars in what follows \citep{VanEck-Jorissen-99}.}
(star symbols in Fig.~\ref{Fig:HR}) are {\em post-mass-transfer}
systems, and as such, may not be directly compared to M giants
and Tc-rich S stars. Nevertheless, the above discussion, stating that
binaries involving Tc-rich S stars and LPVs should have long orbital periods, leads to a
very surprising conclusion: extrinsic, Tc-poor S stars occupy the region of the
$(e - \log P)$ diagram where binary {\em Tc-rich} S stars are expected
(Fig.~\ref{Fig:elogP_M}). This
implies that {\em the mass-transfer episode must have had little
  impact on the orbital elements.} This is quite surprising at first
sight, since either the orbital period must have shorten dramatically
(for the so-called 'case C' RLOF) or it must have increased considerably (for wind
  accretion), according to standard evolutionary prescriptions
\citep[e.g.,][]{Boffin-Jorissen-88}.
This
conclusion is clearly not compatible with the present data, and calls
for alternative mass-transfer prescriptions, 
as it was already known for quite some time. \citet{Jorissen-03},
\citet{Frankowski-2007a} and \citet{Podsia-2007} have proposed new
avenues to explore. For instance, in the context of the 'transient torus' scenario
proposed by  \citet{Frankowski-2007a}, a near constancy of the orbital
period is not unexpected. 

If it is correct that the mass transfer process does not alter much
the orbital period, then the existence of mild barium stars
with rather short orbital periods (i.e., the open triangles to the left of the
solid line in Fig.~\ref{Fig:elogP_panels})
may be explained by the fact that their polluting companion did not evolve far
up the TP-AGB, hence the AGB envelope was not much enriched in s-process elements. This hypothesis is especially appealing since the other
possibility, suggested by \citet{Jorissen-Boffin-92} and \citet{Jorissen-VE-98} -- 
namely, that mild barium stars belong to a more metal-rich
population than strong barium stars, because the s-process
is more   efficient in low-metallicity stars \citep{Goriely-00} -- 
has been dismissed by the detailed abundance
analysis of mild barium stars by \citet{Smiljanic-2007}, who conclude
that there is no obvious metallicity difference between mild and
strong barium stars.

The intriguing absence of s-process enrichment in red symbiotics
\citep[e.g., ][]{Jorissen-03a,Frankowski-2007a}, despite the
fact that they are likely post-mass-transfer systems like barium
stars, may perhaps be understood by invoking the same argument: as
they fall in the same region as the (non-s-process-enriched) post-AGB stars
in the $(e - \log P)$ diagram (Fig.~\ref{Fig:elogP_panels}), it is
possible that their companion did not evolve far up enough on the AGB
to activate dredge-ups and s-process.

\begin{figure}
\includegraphics[width=\columnwidth]{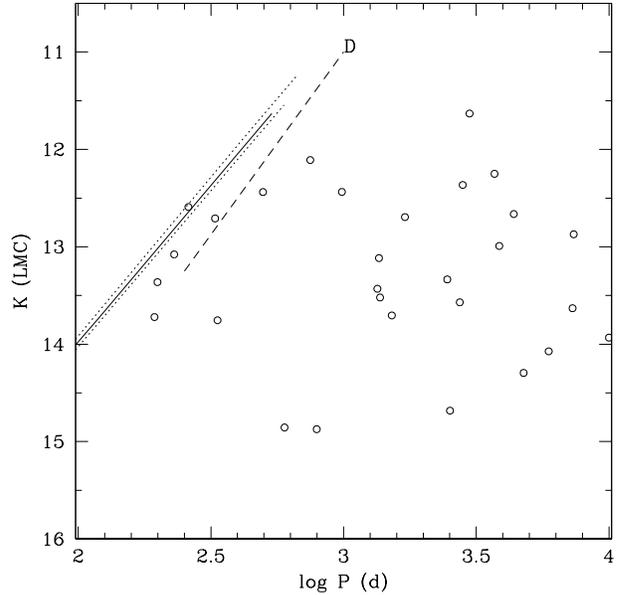}
\caption[]{\label{Fig:MbolP}
The orbital-period as a function of (absolute) $K$ magnitude for
galactic M giants in binary
systems. The ordinate
axis corresponds to the $K$ magnitude that the
M giants would have if they were put at the distance of the LMC (see
text). 
The dashed line labelled 'D' corresponds to Wood's
sequence of long secondary periods. 
The solid and dotted lines correspond to the limit on the
orbital period imposed by the Roche radius (for $M_1 = 1.3$~\Msun\ and $M_2 =
0.6$~\Msun), for the stellar radii obtained from the Stefan-Boltzmann
law applied on the $(V-K, M_{\rm bol})$ pairs defined by the same
lines in Fig.~\ref{Fig:HR1}.   
}
\end{figure}

\subsection{Radius, period and Wood's sequence D in the period --
  luminosity diagram}
\label{Sect:D}

Interestingly enough, our data may be used to shed light on the
much debated nature of the so-called 'sequence D' found by
\citet{Wood-1999} and \citet{Wood-2000} in the period -- luminosity
diagram of LPVs in the Large Magellanic Cloud (LMC). This sequence 
(also called 'long secondary periods') is located to the {\em right}
of the period -- luminosity relationship for Miras pulsating in the
fundamental mode ('sequence C'). Sequence D can thus not be associated with higher
harmonics, which have {\em shorter} periods ('sequences A' and 'B'). A possible relation
between Wood's sequence D and the Roche limit in binary systems (via
ellipsoidal variations or dust obscuration events) was originally
proposed by \citet{Wood-1999} and \citet{Wood-2000}, and more recently
again by \citet{Soszynski-2004b}, \citet{Derekas-2006}
and \citet{Soszynski-2007}.

Fig.~\ref{Fig:MbolP} is a variant of Fig.~\ref{Fig:radius} that makes
it easy to compare our data with the period -- luminosity diagram of
LPVs in the LMC \citep{Wood-1999,Wood-2000}. The ordinate
axis of Fig.~\ref{Fig:MbolP} corresponds to the $K$ magnitude that our
M giants would have if they were at the distance of the LMC, i.e., $K
({\rm LMC}) = M_K + 18.50$, where the \citet{McNamara-2007} LMC distance modulus   
has been adopted. It is quite interesting to notice that Wood's
sequence D does match the upper envelope of
the region occupied by the galactic binary M giants.  Our finding
of a clear relationship between Wood's D sequence and the Roche limit in
binary systems involving M giants 
supports the similar suggestion made by \citet{Soszynski-2004b} and \citet{Soszynski-2007}.

\subsection{Near absence of circular orbits among M~III binaries}
\label{Sect:absence_circular}

\begin{figure}
\includegraphics[width=\columnwidth]{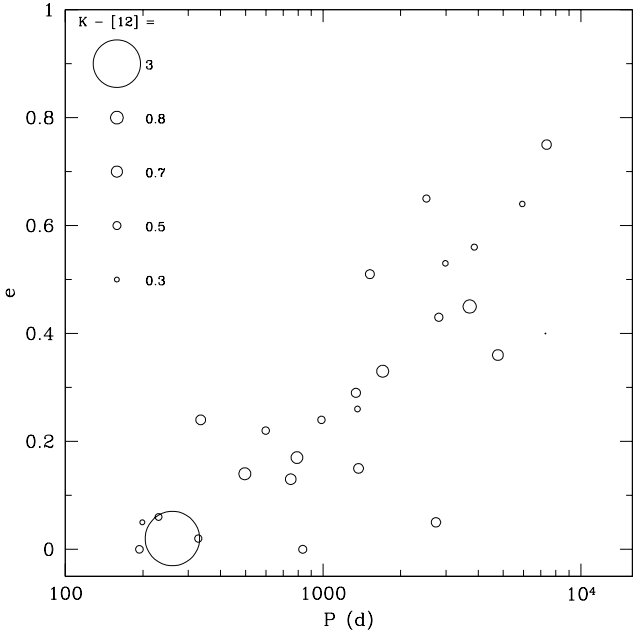}
\caption[]{\label{Fig:elogP_K12}
The $(e - \log P)$ diagram for M giants, where the symbol size is
proportional to the infrared $K - [12]$ index, indicative of the
presence of dust in
the system. The $K$ magnitude is from 2MASS, and $[12] = -2.5
\log(F12/28.3)$ is the magnitude associated with the IRAS 12~$\mu$m flux. $K - [12]$ of the order 0.3 -- 0.5 is expected for non-dusty stars. 
}
\end{figure}

A striking  difference between K and M giants apparent 
on Fig.~\ref{Fig:elogP_M} is the {\em 
lack, among binaries involving M giants, of the many circular systems
at the short end of the period range that are
observed among the K giants}. 
The short-period circular systems observed among K giants result from
tidal effects which circularise the orbit when the giant star is close
to filling its Roche lobe \citep[e.g., ][]{North-92}.  
The difference between K and M giants is surprising, since
both stellar families involve 
stars with deep convective envelopes which should react similarly to
tidal  effects. 
M giants, however, have shorter lifetimes than K giants, because the former suffer from a much more severe wind mass loss, and this difference offers perhaps 
a clue to account for the differences observed at short periods in
their $(e - \log P)$ diagrams. Tidal forces may not have enough time to efficiently operate in the case of M giants. 
Indeed, the many circular systems observed
among post-AGB stars seem to be the circular population missing among the M giants. This hints at mass loss and tidal effects  operating on
similar time scales: shortly after the system has
been circularised, the envelope is lost and the primary star no longer
looks as an M giant but rather as a post-AGB star. 
The very existence of a system like SS~Lep does lend support to the above suggestion.  
In Fig.~\ref{Fig:elogP_K12} investigating a possible correlation
between dust excess and position of the binary M giants in the $(e - \log P)$
diagram, SS~Lep, having a large $K - [12]$ excess,  appears as the big circle along the tidal boundary.

Similarly, it is interesting to note that circular orbits are common
among red symbiotics as well, albeit with somewhat longer periods. 
The implications of this result actually depend on whether the red
symbiotics are pre-mass-transfer systems (i.e., with a main-sequence
companion) or post-mass-transfer systems (i.e., with a WD
companion). This issue has not been solved yet
\citep{Frankowski-2007a}. In the former case, the longer periods are
simply a consequence of the fact that the M giants in red symbiotics are on average of
later types
 (i.e., later than M3) than those in our binary sample
\citep{Belczynski00}. Hence, they cannot be hosted by systems as close
as those hosting earlier giants (Fig.~\ref{Fig:MbolP}). 
The absence of s-process enrichments in those
red symbiotics -- another puzzling issue
\citep{Jorissen-03a,Frankowski-2007a} -- would then be explained naturally as previously for the
post-AGB stars: the M giant has simply not yet reached the TP-AGB phase. 
In the latter case (the companion is a WD), the orbital periods
fall in the short-period tail of the distribution of periods for
post-mass-transfer systems like Tc-poor S stars, as expected, but
then, the absence of s-process enrichment is puzzling unless these
systems had not very evolved AGB progenitors.     

\subsection{The long-period, small-eccentricity gap in the $(e - \log
  P)$ diagram}

To conclude this discussion of the $(e - \log P)$ diagram, 
it must be pointed out that, for  {\it pre}-mass-transfer binaries hosting
main sequence stars, there is a void in this diagram 
at $P > 400$~d, $e < 0.1$. This gap is especially
apparent in the $(e - \log P)$ diagram of G and K dwarfs of the solar
neighbourhood \citep{Duquennoy-Mayor-91}.  
The situation is quite different for  {\em
  post-mass-transfer} systems like barium and Tc-poor S stars.
Figure~1 of \citet{North-00} 
clearly shows that barium dwarfs fall almost exclusively within this gap. 
Similarly, Barium and Tc-poor S stars fill this gap almost completely
(Fig.~\ref{Fig:elogP_panels}). Red symbiotics fill the gap till 1000~d
with circular systems, but as discussed in
Sect.~\ref{Sect:absence_circular}, this may result as well from tidal effects.

There are  in fact a few binaries involving K and M giants (most
notably HD~165374,  with $P = 2741$~d, $e = 0.05$ and a M2III primary)
which fall
in the long-period tail of this gap, as can be seen on
Fig.~\ref{Fig:elogP_panels}. It is very likely   
that these systems are in fact {\it post}-mass-transfer binaries, 
a possibility that
would be worth testing by looking for Ba-like abundance anomalies in
these giants. 

\begin{figure}[!t]
\includegraphics[width=\columnwidth]{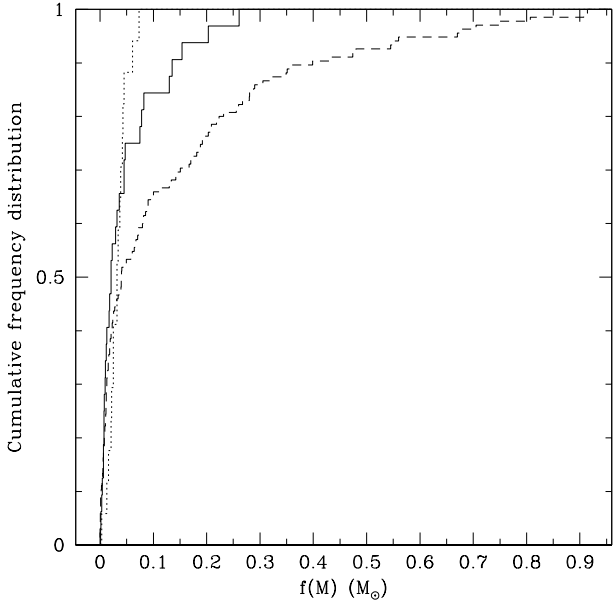}
\caption{\label{Fig:fM}
The cumulative distribution of the mass function for M giants (solid
line), K giants from open clusters \citep[dashed
  line;][]{Mermilliod-2007b} 
and S stars with WD companions \citep[thus excluding
the peculiar systems HD~191589 and HD~332077;][ dotted line]{Jorissen-VE-98}.
}
\end{figure}

An indirect way to confirm the {\it post}-mass-transfer nature of the 
binaries with long periods and small eccentricities is
to look for the presence of WD
companions. \citet{McClure-Woodsworth-90} have convincingly shown that
the cumulative distribution of the mass functions may be used for that
purpose: since WDs have masses spanning a rather narrow range (0.5 --
1.4~\Msun, with a peak around 0.6~\Msun, at least for {\em single}
WDs), the mass functions of binaries involving a giant primary and a WD
secondary may be expected to be much more peaked than in the case of
main-sequence companions, which do not obey a similar constraint on
their masses (except for being less massive than the primary, giant
component). This is illustrated on Fig.~\ref{Fig:fM}, which compares
the cumulative mass-function distributions for M giants, K giants in
open clusters \citep{Mermilliod-2007b} and S stars with WD
companions \citep{Jorissen-VE-98}. The sample of M giants may clearly be split in two subsamples, one with small mass functions matching those of Tc-poor S stars with WD companions, and the other with larger mass functions typical of main-sequence companions, like the remarkable system SS~Lep discussed in this paper ($f(M) = 0.26$~\Msun). 

Conversely, with a mass function of 0.078~\Msun, HD~165374 could certainly host a WD companion of (minimum) mass 0.73~\Msun\ (respectively 0.78~\Msun), 
adopting a mass of 1.5~\Msun\ (respectively 1.7~\Msun) for the M2 giant,
according to its location in the HR diagram of Fig.~\ref{Fig:HR}. 

HD~108907 (4~Dra) is probably a post-mass transfer
system as well, since it hosts a blue, hot
companion, either a cataclysmic variable \citep{Reimers-1988} or more likely a single WD
accreting from the wind of its red giant companion, as in normal
symbiotic systems \citep{Wheatley-2003,Skopal-2005a,Skopal-2005}.
Its position in the $(e -\log P)$ diagram ($e= 0.33, P= 1703$~d) does
not by itself hints at its post-mass-transfer nature, although it
lies just at the border of the region occupied by barium and Tc-poor S
stars, and close to that of the mild barium stars discussed above. 
It does not seem, however, to bear chemical anomalies typical
of barium stars. It may therefore be yet another example of a system 
containing all the necessary ingredients for being a barium star,
but which is not. Again, one may invoke the possibility that the AGB
progenitor did not evolve far enough on the TP-AGB. 

Sect.~\ref{Sect:fM} furthers the considerations of the present subsection by showing that the long-period,
low-eccentricity region of the \elogP\ diagram mostly contains systems
with small mass functions hinting at WD companions. 

\section{A clear signature of mass transfer in the \elogP\ diagram of
  binaries involving K giants in open clusters}
\label{Sect:fM}

\begin{figure}
\includegraphics[width=\columnwidth]{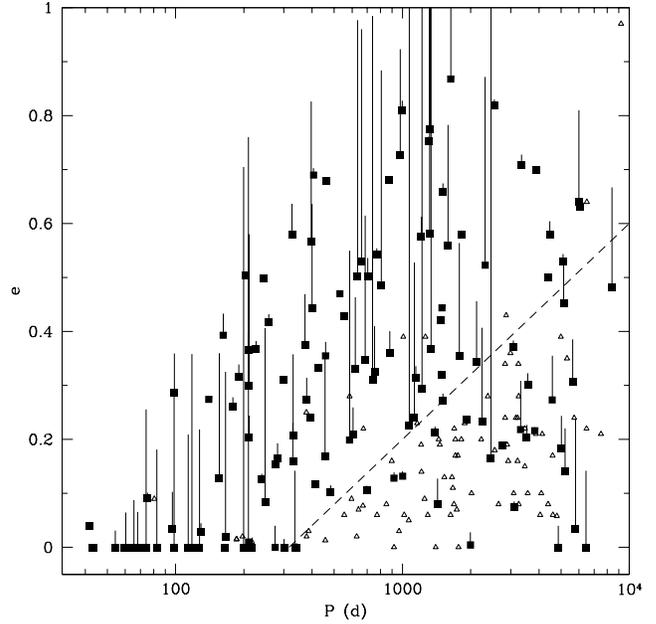}
\caption{\label{Fig:elogP_fM}
The eccentricity--period diagram for spectroscopic binaries
involving K giants in open clusters \citep[filled
  squares;][]{Mermilliod-2007b}, and  
Ba and S stars (open triangles, 
from Jorissen et al.\@ 1998). The vertical segments have lengths
corresponding to mass functions expressed in solar masses.
The diagonal dashed line delineates the region where most of the Ba/S
stars are found.
}
\end{figure}

\begin{figure}[!t]
\includegraphics[width=\columnwidth]{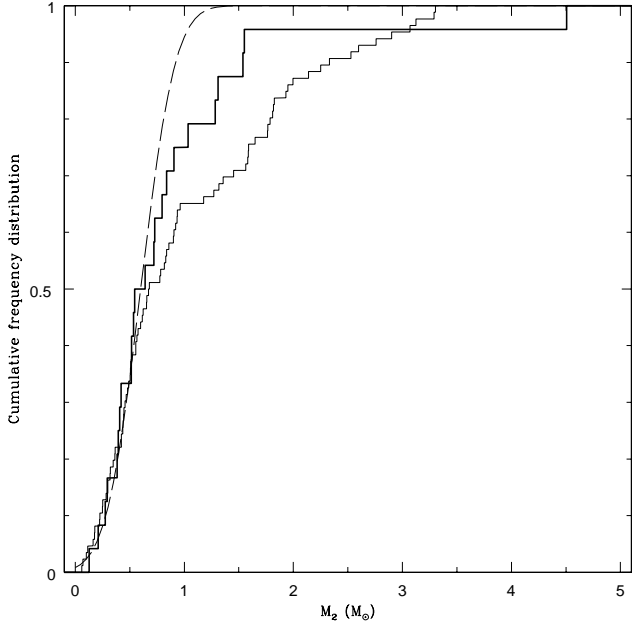}
\caption{\label{Fig:fM_K}
The cumulative distribution of the companion masses for open-cluster K
binaries, assuming an inclination of
90$^\circ$ for all systems (the plotted masses thus represent a
lower bound to the true values): thick solid
line: systems located  in the post-mass-transfer  region
of the \elogP\ diagram, i.e., in the lower-right region below the dashed
line plotted on Fig.~\ref{Fig:elogP_fM}; thin solid line: supposedly non-post-mass-transfer systems (i.e., located above the dashed
line of Fig.~\ref{Fig:elogP_fM});
dashed line: 
for a Gaussian distribution centered on
0.6~\Msun\ with a standard deviation of 0.25~\Msun.
}
\end{figure}

\begin{figure}[!t]
\includegraphics[width=\columnwidth]{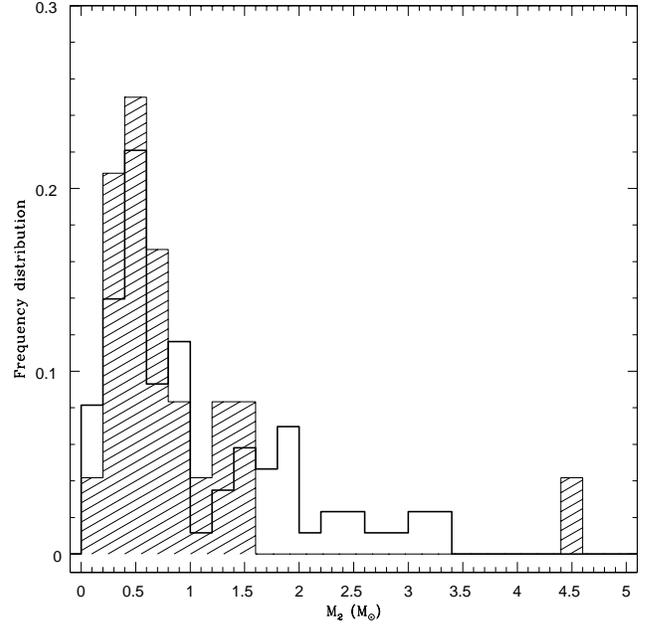}
\caption{\label{Fig:fM_K_histo}
Same as Fig.~\ref{Fig:fM_K}, but in the form of histograms (thick solid line:
systems in the non-post-mass-transfer zone; thin solid line and shaded histogram: post-mass-transfer
systems). 
In the post-mass-transfer region, all companions are less massive than 
1.5~\Msun, in accordance with the expectation for WDs, except for the companion to star 170 in NGC~129, which is a young cluster with a turnoff mass of 5.5~\Msun.
}
\end{figure}

\citet{Mermilliod-2007b} provided an extensive set of orbital elements
for 135 spectroscopic binaries with K giant primaries belonging to open clusters.   
The major asset of this sample, over one involving field K giants as
available from the Ninth Catalogue of Spectroscopic Binary Orbits
\citep{Pourbaix-2004}, is that the mass of the K giant is known: with
a good accuracy, it may be identified with the cluster main-sequence
turnoff mass (it is therefore essential to keep only cluster members
in our study). A lower bound on the mass of the companion may then be
easily obtained from the mass function $f(M)$ by setting $\sin i = 1$.
The distribution of the companion masses in different regions of the
\elogP\ diagram may then be compared. An obvious way to divide the
\elogP\ diagram is through the line of equation $e = 0.4 \log P -1 $
[connecting the $(e, \log P)$ pairs (0, 2.5) and (0.6, 4)] under which
lie most of the barium and Tc-poor S stars (Fig.~\ref{Fig:elogP_fM}). The results of this procedure is shown in Figs.~\ref{Fig:fM_K} and \ref{Fig:fM_K_histo}.

 It is quite clear already from Fig.~\ref{Fig:elogP_fM} that the systems
 involving K giants and located in the same region of the
 \elogP\ diagram as the barium and Tc-poor S stars have a
 mass-function distribution which is significantly different from the
 distribution of the full sample. This is confirmed by Fig.~\ref{Fig:fM_K}. Moreover, these systems lying in
 the post-mass-transfer region have  companions masses below 1.5~\Msun\ (with one exception discussed below), and peaking at 0.5~\Msun, as expected for WD companions 
 (Figs.~\ref{Fig:fM_K} and \ref{Fig:fM_K_histo}). Since most of the clusters making up this supposedly post-mass-transfer sample have turn-off masses well {\it above} 1.5~\Msun, the 1.5~\Msun\ threshold observed  for the companion masses cannot be due to a selection effect.   
 Star~170 (with $P = 2457\pm5$~d, $e = 0.16\pm0.01$ and $f(M) =
 0.91\pm0.04$~\Msun) in NGC~129 \citep[a young cluster with a
 turnoff mass of 5.5~\Msun;][]{Mermilliod-1987} is the
 only outlier, with a $> 4.5$~\Msun\ companion in a  KII + BV pair. 
 
 With a maximum vertical distance of 0.25 between the cumulative frequency distributions of the two samples displayed in  Fig.~\ref{Fig:fM_K} (containing 29 and 106 stars), the  
Kolmogorov-Smirnov statistics leads to reject the null hypothesis that the two samples are extracted from the same parent population with a significance level of 10\% (Performing the same comparison without star 170 in NGC~129 
yields a maximum difference of 0.29 and a significance level of about 4\%). 

\section{Conclusions}

The main results of this paper are the following:\\
1. The \elogP\ diagram of pre-mass transfer binaries like most of the K binaries in open clusters and the binary M giants is bounded by an upper left envelope corresponding to a constant periastron distance. This distance corresponds to the maximum radius reached at a given spectral type, as shown from the similar boundary observed in the radius -- orbital period diagram. Systems lying along that boundary often host a giant (nearly) filling its Roche  lobe, and exhibiting either ellipsoidal variations (HD~9053, HD~190658) or mass-transfer signatures (like an inflated companion; SS~Lep). The mass-transfer binary SS~Lep (MIII + A), with its circumbinary disc most probably fed by non-conservative RLOF, represents a transition case between binary M giants and genuine post-AGB stars.\\
2. Systems similar to SS~Lep and HD~172481, hosting an accreting star mimicking a (super)giant star, should be searched for among post-AGB systems lying along the upper envelope of the \elogP\ diagram. We suggest that the family of binary post-AGB stars might include systems in an active phase of mass transfer, where the accreting component has swollen to giant dimensions, thus mimicking a genuine post-AGB star. \\
3. The so-called Wood's sequence D in the period -- luminosity diagram of LMC LPVs is closely linked to the upper envelope of the \elogP\ diagram for M giants.\\
4. The post-mass transfer systems like barium and Tc-poor S stars have orbital elements very similar to the pre-mass-transfer systems (binary M giants), suggesting that the mass transfer does not alter them much, in line with the new transient-torus scenario for mass transfer proposed by \citet{Frankowski-2007a}.\\
5. The lower right region of the \elogP\ diagram, as defined by the location of the barium and Tc-poor S stars, host mainly post-mass-transfer systems, also among K binaries in open clusters. This is very clearly seen from the fact that these systems have low mass functions consistent with WD companions. Whether these giants bear chemical signatures of mass transfer as do barium stars remains to be checked.

\acknowledgements{We thank J.-Cl. Mermilliod for sending us his orbital data for K giants in open clusters in advance of publication. This work has been partly funded by an {\it Action de recherche concert\'ee (ARC)} from the {\it Direction g\'en\'erale de l'Enseignement non obligatoire et de la Recherche scientifique -- Direction de la recherche scientifique -- Communaut\'e fran\c caise de Belgique.}}
\bibliographystyle{apj} 
\bibliography{ajorisse_articles} 

\begin{thebibliography}{}

\bibitem[\protect\astroncite{{Adelman}}{2007}]{Adelman-2007}
{Adelman} S.~J. 2007, The Journal of Astronomical Data, 13, 3

\bibitem[\protect\astroncite{{Alcock} et~al.}{1998}]{Alcock-1998}
{Alcock} C., {Allsman} R.~A., {Alves} D.~R., {Axelrod} T.~S., {Becker} A.,
  {Bennett} D.~P., {Cook} K.~H., {Freeman} K.~C., {Griest} K., {Lawson} W.~A.,
  {Lehner} M.~J., {Marshall} S.~L., {Minniti} D., {Peterson} B.~A., {Pollard}
  K.~R., {Pratt} M.~R., {Quinn} P.~J., {Rodgers} A.~W., {Sutherland} W.,
  {Tomaney} A., {Welch} D.~L. 1998, \aj, 115, 1921

\bibitem[\protect\astroncite{{Allen}}{1982}]{Allen-82}
{Allen} D.~A. 1982,
\newblock in M. {Friedjung}, R. {Viotti} (eds.), The Nature of Symbiotic Stars
  (IAU Coll. 70), Reidel, Dordrecht, ~27

\bibitem[\protect\astroncite{{Alvarez} et~al.}{2001}]{Alvarez-2001}
{Alvarez} R., {Jorissen} A., {Plez} B., {Gillet} D., {Fokin} A., {Dedecker} M.
  2001, \aap, 379, 305

\bibitem[\protect\astroncite{{Artymowicz} et~al.}{1991}]{Artymowicz91}
{Artymowicz} P., {Clarke} C.~J., {Lubow} S.~H., {Pringle} J.~E. 1991, \apjl,
  370, L35

\bibitem[\protect\astroncite{{Ayres}}{2005}]{Ayres-2005}
{Ayres} T.~R. 2005, \apj, 618, 493

\bibitem[\protect\astroncite{{Bakker} et~al.}{1996}]{Bakker-1996}
{Bakker} E.~J., {Lamers} H.~J.~G.~L.~M., {Waters} L.~B.~F.~M., {Waelkens} C.,
  {Trams} N.~R., {Van Winckel} H. 1996, \aap, 307, 869

\bibitem[\protect\astroncite{{Barnes} \& {Evans}}{1976}]{Barnes-Evans-1976}
{Barnes} T.~G., {Evans} D.~S. 1976, \mnras, 174, 489

\bibitem[\protect\astroncite{{Belczy{\' n}ski} et~al.}{2000}]{Belczynski00}
{Belczy{\' n}ski} K., {Miko{\l}ajewska} J., {Munari} U., {Ivison} R.~J.,
  {Friedjung} M. 2000, A\&AS, 146, 407

\bibitem[\protect\astroncite{{Benz} \& {Mayor}}{1981}]{Benz-1981}
{Benz} W., {Mayor} M. 1981, \aap, 93, 235

\bibitem[\protect\astroncite{{Bessell} et~al.}{1998}]{Bessell-98}
{Bessell} M.~S., {Castelli} F., {Plez} B. 1998, \aap, 333, 231

\bibitem[\protect\astroncite{{Bessell} \& {Wood}}{1984}]{Bessell-Wood-84}
{Bessell} M.~S., {Wood} P.~R. 1984, PASP, 96, 247

\bibitem[\protect\astroncite{{Bl\"ocker}}{1995}]{Bloecker-95}
{Bl\"ocker} T. 1995, \aap, 297, 727

\bibitem[\protect\astroncite{{Boffin} \& {Jorissen}}{1988}]{Boffin-Jorissen-88}
{Boffin} H. M.~J., {Jorissen} A. 1988, \aap, 205, 155

\bibitem[\protect\astroncite{{Charbonnel} et~al.}{1996}]{Charbonnel-1996}
{Charbonnel} C., {Meynet} G., {Maeder} A., {Schaerer} D. 1996, A\&AS, 115, 339

\bibitem[\protect\astroncite{{Cummings} et~al.}{1999}]{Cummings-1999}
{Cummings} I.~N., {Hearnshaw} J.~B., {Kilmartin} P.~M., {Gilmore} A.~C. 1999,
\newblock in J.~B. {Hearnshaw}, C.~D. {Scarfe} (eds.), IAU Colloq. 170: Precise
  Stellar Radial Velocities, Vol. 185 of {\em Astronomical Society of the
  Pacific Conference Series\/}, p.~204

\bibitem[\protect\astroncite{{de Ruyter} et~al.}{2006}]{DeRuyter-2006}
{de Ruyter} S., {van Winckel} H., {Maas} T., {Lloyd Evans} T., {Waters}
  L.~B.~F.~M., {Dejonghe} H. 2006, \aap, 448, 641

\bibitem[\protect\astroncite{{Derekas} et~al.}{2006}]{Derekas-2006}
{Derekas} A., {Kiss} L.~L., {Bedding} T.~R., {Kjeldsen} H., {Lah} P.,
  {Szab{\'o}} G.~M. 2006, \apjl, 650, L55

\bibitem[\protect\astroncite{{Dermine} et~al.}{2009}]{Dermine-Jorissen-2008}
{Dermine} T., {Jorissen} A., {Siess} L., {Frankowski} A. 2009, \aap, submitted

\bibitem[\protect\astroncite{{Dumm} \& {Schild}}{1998}]{DummSchild98}
{Dumm} T., {Schild} H. 1998, New Astr., 3, 137

\bibitem[\protect\astroncite{{Duquennoy} \& {Mayor}}{1991}]{Duquennoy-Mayor-91}
{Duquennoy} A., {Mayor} M. 1991, \aap, 248, 485

\bibitem[\protect\astroncite{{Duquennoy} et~al.}{1992}]{Duquennoy-92}
{Duquennoy} A., {Mayor} M., {Mermilliod} J. 1992,
\newblock in A. {Duquennoy}, M. {Mayor} (eds.), Binaries as Tracers of Stellar
  Formation, Cambridge University Press, Cambridge, ~52

\bibitem[\protect\astroncite{{Edvardsson} et~al.}{1993}]{Edvardsson-93}
{Edvardsson} B., {Andersen} J., {Gustafsson} B., {Lambert} D.~L., {Nissen}
  P.~E., {Tomkin} J. 1993, \aap, 275, 101

\bibitem[\protect\astroncite{{Eggleton}}{2006}]{Eggleton-2006}
{Eggleton} P. 2006,
\newblock Evolutionary Processes in Binary and Multiple Stars,
\newblock Cambridge University Press

\bibitem[\protect\astroncite{{Eggleton}}{1983}]{Eggleton-83}
{Eggleton} P.~P. 1983, \apj, 268, 368

\bibitem[\protect\astroncite{ESA}{1997}]{ESA-1997}
ESA 1997,
\newblock The Hipparcos and Tycho Catalogues,
\newblock ESA

\bibitem[\protect\astroncite{{Famaey} et~al.}{2005}]{Famaey-2005}
{Famaey} B., {Jorissen} A., {Luri} X., {Mayor} M., {Udry} S., {Dejonghe} H.,
  {Turon} C. 2005, \aap, 430, 165

\bibitem[\protect\astroncite{{Famaey} et~al.}{2009}]{Famaey-2008:b}
{Famaey} B., {Pourbaix} D., {Jorissen} A., {Frankowski} M., {Van Eck} S.,
  {Mayor} M., {Udry} S. 2009, \aap, this volume (Paper I)

\bibitem[\protect\astroncite{{Fekel} et~al.}{2007}]{Fekel-2007}
{Fekel} F.~C., {Hinkle} K.~H., {Joyce} R.~R., {Wood} P.~R., {Lebzelter} T.
  2007, \aj, 133, 17

\bibitem[\protect\astroncite{{Frankowski}}{2003}]{Frankowski-2003}
{Frankowski} A. 2003, \aap, 406, 265

\bibitem[\protect\astroncite{{Frankowski}}{2008}]{Frankowski-2008:b}
{Frankowski} A. 2008,
\newblock in R. {Corradi}, A. {Manchado}, N. {Soker} (eds.), Asymmetrical
  Planetary Nebulae IV, electronic publication available at
  http://www.iac.es/project/apn4/

\bibitem[\protect\astroncite{{Frankowski} et~al.}{2009}]{Frankowski-2008}
{Frankowski} A., {Famaey} B., {Van Eck} S., {Jorissen} A. 2009, \aap, this
  volume (Paper II)

\bibitem[\protect\astroncite{{Frankowski} \&
  {Jorissen}}{2007}]{Frankowski-2007a}
{Frankowski} A., {Jorissen} A. 2007, Baltic Astronomy, 16, 104

\bibitem[\protect\astroncite{{Goldin} \& {Makarov}}{2007}]{Goldin-Makarov-2007}
{Goldin} A., {Makarov} V.~V. 2007, \apjs, 173, 137

\bibitem[\protect\astroncite{{Goriely} \& {Mowlavi}}{2000}]{Goriely-00}
{Goriely} S., {Mowlavi} N. 2000, \aap, 362, 599

\bibitem[\protect\astroncite{{Herman} \& {Habing}}{1985}]{HermanHabing85}
{Herman} J., {Habing} H.~J. 1985, Phys. Rep., 124, 255

\bibitem[\protect\astroncite{{Hinkle} et~al.}{1997}]{Hinkle-1997}
{Hinkle} K.~H., {Lebzelter} T., {Scharlach} W.~W.~G. 1997, \aj, 114, 2686

\bibitem[\protect\astroncite{{Humphreys}}{1976}]{Humphreys-1976}
{Humphreys} R.~M. 1976, \apj, 206, 122

\bibitem[\protect\astroncite{{H\"unsch} \& {Schroeder}}{1996}]{Hunsch-1996}
{H\"unsch} M., {Schroeder} K.-P. 1996, \aap, 309, L51

\bibitem[\protect\astroncite{{Hut}}{1981}]{Hut81}
{Hut} P. 1981, \aap, 99, 126

\bibitem[\protect\astroncite{{Iben} \& {Tutukov}}{1996}]{Iben-Tutukov-1996}
{Iben} I.~J., {Tutukov} A.~V. 1996, \apjs, 105, 145

\bibitem[\protect\astroncite{{Jancart} et~al.}{2005}]{Jancart-2005}
{Jancart} S., {Jorissen} A., {Babusiaux} C., {Pourbaix} D. 2005, \aap, 442, 365

\bibitem[\protect\astroncite{{Jorissen}}{2003a}]{Jorissen-03}
{Jorissen} A. 2003a,
\newblock in H. {Habing}, H. {Olofsson} (eds.), Asymptotic Giant Branch Stars,
  Springer Verlag, New York, p.~461

\bibitem[\protect\astroncite{{Jorissen}}{2003b}]{Jorissen-03a}
{Jorissen} A. 2003b,
\newblock in R.~L.~M. {Corradi}, J. {Miko\l ajewska}, T.~J. {Mahoney} (eds.),
  Symbiotic stars probing stellar evolution, Astron. Soc. Pacific Conf. Ser.
  Vol. 303, San Francisco, ~25

\bibitem[\protect\astroncite{{Jorissen} \& {Boffin}}{1992}]{Jorissen-Boffin-92}
{Jorissen} A., {Boffin} H.~M.~J. 1992,
\newblock in A. {Duquennoy}, M. {Mayor} (eds.), Binaries as Tracers of Stellar
  Formation, Cambridge University Press, Cambridge,  110

\bibitem[\protect\astroncite{{Jorissen} et~al.}{1995}]{Jorissen-95}
{Jorissen} A., {Hennen} O., {Mayor} M., {Bruch} A., {Sterken} C. 1995, \aap,
  301, 707

\bibitem[\protect\astroncite{{Jorissen} \& {Mayor}}{1992}]{Jorissen-Mayor-92}
{Jorissen} A., {Mayor} M. 1992, \aap, 260, 115

\bibitem[\protect\astroncite{{Jorissen} \& {Van
  Eck}}{2005}]{Jorissen-VanEck-2005}
{Jorissen} A., {Van Eck} S. 2005,
\newblock in Cosmic Abundances as Records of Stellar Evolution and
  Nucleosynthesis, ASP Conf. Ser. Vol. 336, San Francisco,  207

\bibitem[\protect\astroncite{{Jorissen} et~al.}{1998}]{Jorissen-VE-98}
{Jorissen} A., {Van Eck} S., {Mayor} M., {Udry} S. 1998, \aap, 332, 877

\bibitem[\protect\astroncite{{Jorissen} et~al.}{2005}]{Jorissen-Zacs-2005}
{Jorissen} A., {Za{\v c}s} L., {Udry} S., {Lindgren} H., {Musaev} F.~A. 2005,
  \aap, 441, 1135

\bibitem[\protect\astroncite{{Jura} et~al.}{2002}]{Jura-2002}
{Jura} M., {Chen} C., {Plavchan} P. 2002, \apj, 569, 964

\bibitem[\protect\astroncite{{Jura} et~al.}{2001}]{Jura-2001}
{Jura} M., {Webb} R.~A., {Kahane} C. 2001, \apjl, 550, L71

\bibitem[\protect\astroncite{{Kippenhahn} \&
  {Meyer-Hofmeister}}{1977}]{Kippenhahn77}
{Kippenhahn} R., {Meyer-Hofmeister} E. 1977, \aap, 54, 539

\bibitem[\protect\astroncite{{Kipper}}{2005}]{Kipper-2005}
{Kipper} T. 2005, Baltic Astronomy, 14, 223

\bibitem[\protect\astroncite{{Lebzelter} et~al.}{2005}]{Lebzelter-2005}
{Lebzelter} T., {Hinkle} K.~H., {Wood} P.~R., {Joyce} R.~R., {Fekel} F.~C.
  2005, \aap, 431, 623

\bibitem[\protect\astroncite{{Lewis} et~al.}{1987}]{Lewis-1987}
{Lewis} B.~M., {Eder} J., {Terzian} Y. 1987, AJ, 94, 1025

\bibitem[\protect\astroncite{{Little} et~al.}{1987}]{Little-87}
{Little} S.~J., {Little-Marenin} I.~R., {Bauer} W.~H. 1987, AJ, 94, 981

\bibitem[\protect\astroncite{{L\"u}}{1991}]{Lu-91}
{L\"u} P.~K. 1991, AJ, 101, 2229

\bibitem[\protect\astroncite{{Lucke} \& {Mayor}}{1982}]{Lucke-Mayor-1982}
{Lucke} P.~B., {Mayor} M. 1982, \aap, 105, 318

\bibitem[\protect\astroncite{{McClure}}{1983}]{McClure-83}
{McClure} R.~D. 1983, ApJ, 268, 264

\bibitem[\protect\astroncite{{McClure} \&
  {Woodsworth}}{1990}]{McClure-Woodsworth-90}
{McClure} R.~D., {Woodsworth} A.~W. 1990, ApJ, 352, 709

\bibitem[\protect\astroncite{{McNamara} et~al.}{2007}]{McNamara-2007}
{McNamara} D.~H., {Clementini} G., {Marconi} M. 2007, \aj, 133, 2752

\bibitem[\protect\astroncite{{Mermilliod} et~al.}{2007}]{Mermilliod-2007b}
{Mermilliod} J.-C., {Andersen} J., {Latham} D.~W., {Mayor} M. 2007, \aap, 473,
  829

\bibitem[\protect\astroncite{{Mermilliod} et~al.}{1987}]{Mermilliod-1987}
{Mermilliod} J.~C., {Mayor} M., {Burki} G. 1987, \aaps, 70, 389

\bibitem[\protect\astroncite{{Miko\l ajewska}}{2003}]{Mikolajewska-03}
{Miko\l ajewska} J. 2003,
\newblock in R.~L.~M. {Corradi}, J. {Miko\l ajewska}, T.~J. {Mahoney} (eds.),
  Symbiotic stars probing stellar evolution, Astron. Soc. Pacific Conf. Ser.
  Vol. 303, San Francisco, ~9

\bibitem[\protect\astroncite{{Nelson} \& {Eggleton}}{2001}]{Nelson-2001}
{Nelson} C.~A., {Eggleton} P.~P. 2001, \apj, 552, 664

\bibitem[\protect\astroncite{{North} \& {Duquennoy}}{1992}]{North-92}
{North} P., {Duquennoy} A. 1992,
\newblock in A. {Duquennoy}, M. {Mayor} (eds.), Binaries as Tracers of Stellar
  Formation, Cambridge University Press, Cambridge,  202

\bibitem[\protect\astroncite{{North} et~al.}{2000}]{North-00}
{North} P., {Jorissen} A., {Mayor} M. 2000,
\newblock in R.~F. {Wing} (ed.), The Carbon Star Phenomenon (IAU Symp. 177),
  Kluwer, Dordrecht,  269

\bibitem[\protect\astroncite{{Otero}}{2007}]{Otero-2007}
{Otero} S.~A. 2007, Open European Journal on Variable Stars, 72, 1

\bibitem[\protect\astroncite{{Paczy\'nski} \&
  {Rudak}}{1980}]{Paczynski-Rudak-80}
{Paczy\'nski} B., {Rudak} B. 1980, \aap, 82, 349

\bibitem[\protect\astroncite{{Pastetter} \&
  {Ritter}}{1989}]{Pastetter-Ritter-89}
{Pastetter} L., {Ritter} H. 1989, \aap, 214, 186

\bibitem[\protect\astroncite{{Podsiadlowski} \& {Mohamed}}{2007}]{Podsia-2007}
{Podsiadlowski} P., {Mohamed} S. 2007, Baltic Astronomy, 16, 26

\bibitem[\protect\astroncite{{Pourbaix}}{2004}]{Pourbaix-2004}
{Pourbaix} D. 2004,
\newblock in R.~W. {Hilditch}, H. {Hensberge}, K. {Pavlovski} (eds.),
  Spectroscopically and Spatially Resolving the Components of Close Binary
  Stars, San Francisco: Astron. Soc. Pacific Conf. Ser. 318, p.~132

\bibitem[\protect\astroncite{{Pourbaix} et~al.}{2004}]{Pourbaix-04a}
{Pourbaix} D., {Tokovinin} A.~A., {Batten} A.~H., {Fekel} F.~C., {Hartkopf}
  W.~I., {Levato} H., {Morrell} N.~I., {Torres} G., {Udry} S. 2004, \aap, 424,
  727

\bibitem[\protect\astroncite{{Ragland} et~al.}{2006}]{Ragland-2006}
{Ragland} S., {Traub} W.~A., {Berger} J.-P., {Danchi} W.~C., {Monnier} J.~D.,
  {Willson} L.~A., {Carleton} N.~P., {Lacasse} M.~G., {Millan-Gabet} R.,
  {Pedretti} E., {Schloerb} F.~P., {Cotton} W.~D., {Townes} C.~H., {Brewer} M.,
  {Haguenauer} P., {Kern} P., {Labeye} P., {Malbet} F., {Malin} D., {Pearlman}
  M., {Perraut} K., {Souccar} K., {Wallace} G. 2006, \apj, 652, 650

\bibitem[\protect\astroncite{{Reimers} et~al.}{1988}]{Reimers-1988}
{Reimers} D., {Griffin} R.~F., {Brown} A. 1988, \aap, 193, 180

\bibitem[\protect\astroncite{{Reimers} et~al.}{1996}]{Reimers-1996}
{Reimers} D., {Huensch} M., {Schmitt} J.~H.~M.~M., {Toussaint} F. 1996, \aap,
  310, 813

\bibitem[\protect\astroncite{{Reyniers} \& {Van Winckel}}{2001}]{Reyniers-2001}
{Reyniers} M., {Van Winckel} H. 2001, \aap, 365, 465

\bibitem[\protect\astroncite{{Richichi} \& {Percheron}}{2005}]{Richichi-2005b}
{Richichi} A., {Percheron} I. 2005, \aap, 434, 1201

\bibitem[\protect\astroncite{{Richichi} et~al.}{2005}]{Richichi-2005}
{Richichi} A., {Percheron} I., {Khristoforova} M. 2005, \aap, 431, 773

\bibitem[\protect\astroncite{{Royer} et~al.}{2002}]{Royer-2002}
{Royer} F., {Grenier} S., {Baylac} M.-O., {G{\'o}mez} A.~E., {Zorec} J. 2002,
  \aap, 393, 897

\bibitem[\protect\astroncite{{Samus}}{1997}]{Samus-1997}
{Samus} N.~N. 1997, Informational Bulletin on Variable Stars, 4501, 1

\bibitem[\protect\astroncite{{Schaller} et~al.}{1992}]{Schaller-1992}
{Schaller} G., {Schaerer} D., {Meynet} G., {Maeder} A. 1992, A\&AS, 96, 269

\bibitem[\protect\astroncite{{Schmid} \& {Schild}}{2002}]{Schmid-Schild-2002}
{Schmid} H.~M., {Schild} H. 2002, \aap, 395, 117

\bibitem[\protect\astroncite{{Schwarz} et~al.}{1995}]{Schwarz-1995}
{Schwarz} H.~E., {Nyman} L.-A., {Seaquist} E.~R., {Ivison} R.~J. 1995, \aap,
  303, 833

\bibitem[\protect\astroncite{{Sivarani} et~al.}{1999}]{Sivarani-1999}
{Sivarani} T., {Parthasarathy} M., {Garc{\'{\i}}a-Lario} P., {Manchado} A.,
  {Pottasch} S.~R. 1999, \aaps, 137, 505

\bibitem[\protect\astroncite{{Skopal}}{2005a}]{Skopal-2005a}
{Skopal} A. 2005a,
\newblock in J.-M. {Hameury}, J.-P. {Lasota} (eds.), The Astrophysics of
  Cataclysmic Variables and Related Objects, Astron. Soc. Pac. Conf. Ser.
  (Vol.330), San Francisco,  463

\bibitem[\protect\astroncite{{Skopal}}{2005b}]{Skopal-2005}
{Skopal} A. 2005b, \aap, 440, 995

\bibitem[\protect\astroncite{{Smiljanic} et~al.}{2007}]{Smiljanic-2007}
{Smiljanic} R., {Porto de Mello} G.~F., {da Silva} L. 2007, \aap, 468, 679

\bibitem[\protect\astroncite{{Soker}}{2000}]{Soker00}
{Soker} N. 2000, \aap, 357, 557

\bibitem[\protect\astroncite{{Soszy{\'n}ski}}{2007}]{Soszynski-2007}
{Soszy{\'n}ski} I. 2007, \apj, 660, 1486

\bibitem[\protect\astroncite{{Soszy\'nski} et~al.}{2004}]{Soszynski-2004b}
{Soszy\'nski} I., {Udalski} A., {Kubiak} M., {Szyma\'nski} M.~K.,
  {Pietrzy\'nski} G., {\.{Z}ebru\'n} K., {Szewczyk} O., {Wyrzykowski} {\L}.,
  {Dziembowski} W.~A. 2004, Acta Astronomica, 54, 347

\bibitem[\protect\astroncite{{Tomkin} et~al.}{1989}]{Tomkin-89}
{Tomkin} J., {Lambert} D.~L., {Edvardsson} B., {Gustafsson} B., {Nissen} P.~E.
  1989, \aap, 219, L15

\bibitem[\protect\astroncite{{Torres} et~al.}{2002}]{Torres-2002}
{Torres} G., {Neuh{\"a}user} R., {Guenther} E.~W. 2002, \aj, 123, 1701

\bibitem[\protect\astroncite{{Udry} et~al.}{1997}]{Udry-1997}
{Udry} S., {Mayor} M., {Andersen} J., {Crifo} F., {Grenon} M., {Imbert} M.,
  {Lindegren} H., {Maurice} E., {Nordstroem} B., {Pernier} B., {Prevot} L.,
  {Traversa} G., {Turon} C. 1997,
\newblock in M. {Perryman} (ed.), Hipparcos - Venice '97 (ESA SP-402), ESA,
  Noordwijk, p.~693

\bibitem[\protect\astroncite{{Van Eck} \&
  {Jorissen}}{1999}]{VanEck-Jorissen-99}
{Van Eck} S., {Jorissen} A. 1999, \aap, 345, 127

\bibitem[\protect\astroncite{{Van Eck} et~al.}{1998}]{VanEck-98}
{Van Eck} S., {Jorissen} A., {Udry} S., {Mayor} M., {Pernier} B. 1998, \aap,
  329, 971

\bibitem[\protect\astroncite{{Van Winckel}}{2003}]{VanWinckel-03}
{Van Winckel} H. 2003, \araa, 41, 391

\bibitem[\protect\astroncite{{Van Winckel}}{2007}]{VanWinckel-2007}
{Van Winckel} H. 2007, Baltic Astronomy, 16, 112

\bibitem[\protect\astroncite{{Verhoelst} et~al.}{2007}]{Verhoelst-2007}
{Verhoelst} T., {van Aarle} E., {Acke} B. 2007, \aap, 470, L21

\bibitem[\protect\astroncite{{Welty} \& {Wade}}{1995}]{Welty-1995}
{Welty} A.~D., {Wade} R.~A. 1995, \aj, 109, 326

\bibitem[\protect\astroncite{{Wheatley} et~al.}{2003}]{Wheatley-2003}
{Wheatley} P.~J., {Mukai} K., {de Martino} D. 2003, \mnras, 346, 855

\bibitem[\protect\astroncite{{Whitelock} \& {Marang}}{2001}]{Whitelock-2001}
{Whitelock} P., {Marang} F. 2001, \mnras, 323, L13

\bibitem[\protect\astroncite{{Whitelock}}{1987}]{Whitelock87}
{Whitelock} P.~A. 1987, PASP, 99, 573

\bibitem[\protect\astroncite{{Wood}}{2000}]{Wood-2000}
{Wood} P.~R. 2000, \pasa, 17, 18

\bibitem[\protect\astroncite{{Wood} et~al.}{1999}]{Wood-1999}
{Wood} P.~R., {Alcock} C., {Allsman} R.~A., {Alves} D., {Axelrod} T.~S.,
  {Becker} A.~C., {Bennett} D.~P., {Cook} K.~H., {Drake} A.~J., {Freeman}
  K.~C., {Griest} K., {King} L.~J., {Lehner} M.~J., {Marshall} S.~L., {Minniti}
  D., {Peterson} B.~A., {Pratt} M.~R., {Quinn} P.~J., {Stubbs} C.~W.,
  {Sutherland} W., {Tomaney} A., {Vandehei} T., {Welch} D.~L. 1999,
\newblock in Asymptotic Giant Branch Stars (IAU Symp. 191),  151

\bibitem[\protect\astroncite{{Zahn}}{1977}]{Zahn-1977}
{Zahn} J.-P. 1977, \aap, 57, 383

\bibitem[\protect\astroncite{{Zamanov} et~al.}{2006}]{Zamanov-2006}
{Zamanov} R.~K., {Bode} M.~F., {Melo} C.~H.~F., {Porter} J., {Gomboc} A.,
  {Konstantinova-Antova} R. 2006, \mnras, 365, 1215

\end{thebibliography}

\appendix
\section*{Appendix. Dismissing the binary nature of L2 Pup}

The intringuing finding by
\citet{Goldin-Makarov-2007} of a very short period of 141~d in
the Hipparcos Intermediate Astrometric Data  for the M5III
semiregular variable star HIP~34922 (=HD~56096 = L$^2$~Pup) deserves 
a specific discussion.
This 141~d period is clearly
related to pulsation, since that periodicity is found in both the
light  and  radial-velocity curves
\citep{Cummings-1999,Lebzelter-2005}. The light curve has a large
amplitude (about 2.5~mag in the visual), which cannot therefore be
attributed to ellipsoidal variations. According to
\citet{Lebzelter-2005}, the 141~d period moreover falls  on sequence C
of the period-luminosity diagram of \citet{Wood-2000}, which
corresponds to the fundamental pulsation mode of long-period variables
(see Sect.~\ref{Sect:D}). Assuming that the 141~d period corresponds
as well to an orbital motion (in the case of a tidal lock between
orbital motion and pulsations) leads, however, to major
inconsistencies. The absolute semi-major axis for the photocentre of
the system, inferred from Goldin \& Makarov's angular value $a_0 =
9.5$~mas and from the parallax of $\varpi = 16.5$~mas, is 0.58~AU,
which may be considered as the absolute semi-major axis of the orbit
of the giant if one assumes that the companion contributes no light to
the system. According to \citet{DummSchild98}, the minimum radius of
an M5 giant is 60~\Rsun\ \citep[but the  radius predicted by Dumm \&
  Schild from the Barnes-Evans relationship for L$^2$ Pup is more like
  125~\Rsun;][]{Barnes-Evans-1976}. The giant could just fit in the
binary system, filling its Roche lobe at periastron, assuming a
reasonable mass ratio of 1.5 and implying a semi-major axis $A =
1.45$~AU for the relative orbit. But then, according to Kepler's third
law, the total mass of the system would be 20.4~\Msun, or
12.2~\Msun\ for the giant and 8.2~\Msun\ for its companion, which is
inconsistent with the M5III classification of the former. We thus
conclude that, although the Hipparcos data are clearly inconsistent
with a single point source (with residuals from the single-star
solution amounting up to 10~mas), Goldin \& Makarov's astrometric
orbit is impossible to reconcile with current knowledge about this
system. It could well be that the Hipparcos data have been confused by
the asymmetric surface brightness typical of late-type stars
\citep{Ragland-2006}, especially so for a star as extended as L$^2$
Pup having a (predicted) apparent angular radius of 9.6~mas
\citep{DummSchild98}. A further argument {\em against} the binary
nature of L$^2$~Pup is the detection of all three masers SiO, H$_2$O
and OH \citep{Jura-2002}, since a binary companion, when orbiting in
the layers where the maser activity should originate, prevents its
operation \citep{HermanHabing85,Lewis-1987,Schwarz-1995}.

\end{document}